\documentclass{article}

\usepackage{lineno,hyperref}
\usepackage{colortbl}
\usepackage{amsmath,amssymb,amsfonts}
\usepackage{algorithmic}
\usepackage{graphicx}
\usepackage{float}
\usepackage{setspace}
\usepackage{subfigure,dcolumn}
\usepackage[T2A,T1]{fontenc}
\usepackage[russian,english]{babel}
\usepackage{todonotes}
\usepackage{multirow}
\usepackage{listings}
\usepackage{appendix}
\usepackage{url}
\usepackage{lineno}
\usepackage[numbers,sort&compress]{natbib}

\usepackage{authblk}
\title{Expected geoneutrino signal at JUNO using local integrated 3-D refined crustal model}

\author[1]{Ran HAN \thanks{corresponding author: hanran@ncepu.edu.cn}}                  
\author[2]{ZhiWei LI\thanks{corresponding author: zwli@whigg.ac.cn }}          
\author[3]{Ruohan GAO}
\author[4]{Ya XU}
\author[5]{Yufei XI}
\author[6]{Guangzheng JIANG}
\author[7]{Andong WANG}
\author[2]{Yaping CHENG}
\author[3]{Yao SUN} 
\author[1]{Jie PANG}
\author[3]{Qi HUA}
\author[8]{Liangjian WEN}
\author[8]{Liang ZHAN}
\author[8]{YuFeng LI}

\affil[1]{School of Nuclear Science \& Engineering, North China Electric Power University, Beijing 102206, China.}
\affil[2]{Beijing Institute of Spacecraft Environment Engineering, Beijing 100094, China.}
\affil[3]{China University of Geosciences (Beijing), Beijing, 100083, China.}
\affil[4]{Key Laboratory of Petroleum Resource Research, Institute of Geology and Geophysics, Chinese Academy of Sciences, Beijing, 100029, China.}
\affil[5]{Institute of Hydrogeology and Environmental Geology, Chinese Academy of Geological Sciences, Shijiazhuang, 050061, China.}
\affil[6]{State Key Laboratory for Oil and Gas Reservoir Geology and Exploitation, Chengdu University of Technology, Chengdu, 610059, China.}
\affil[7]{East China University of Technology, Nanchang, 330013, China.}
\affil[8]{Institute of High Energy Physics, Chinese Academy of Sciences, Beijing, 100049, China.}

\begin{document}

\maketitle
%\linenumbers  
\newpage
\begin{abstract}

Geoneutrinos serve as a potent tool for comprehending the Earth's radiogenic power and composition.Although geoneutrinos have been observed in prior experiments, the forthcoming generation of experiments, such as JUNO, will be necessary for fully harnessing their potential.Precise prediction of the crustal contribution is vital for interpreting particle physics measurements in the context of geo-scientific inquiries. Nonetheless, existing models such as JULOC and GIGJ have limitations in accurately forecasting the crustal contribution. This paper introduces JULOC-I, JUNO's novel 3-D integrated crustal model, which employs seismic, gravity, rock sample, and heat flow data to precisely estimate the geoneutrino signal of the lithosphere.The model indicates elevated concentrations of uranium and thorium in southern China, resulting in unexpectedly strong geoneutrino signals.

The accuracy of JULOC-I, coupled with a decade of experimental data, affords JUNO the opportunity to test multiple mantle models.Once operational, JUNO can validate the model's predictions and enhance the precision of mantle measurements.All in all, the improved accuracy of JULOC-I represents a substantial stride towards comprehending the geochemical distribution of the South China crust, offering a valuable tool for investigating the composition and evolution of the Earth through geoneutrinos.

\end{abstract}
\newpage
\section{Introduction}

The collaboration between geologists and physicists over the past 150 years has led to the emergence of various scientific disciplines.Although radioactivity plays a crucial role in the thermal history of the Earth, accurately quantifying the amount of radiogenic heat remains challenging.

Although radioactivity plays a crucial role in the thermal history of the Earth, accurately quantifying the amount of radiogenic heat remains challenging. Geothermal heat, generated within the Earth, influences the overall development and evolution of the planet, as well as controlling crustal and mantle activity, the formation of volcanoes, magmatic processes, earthquakes, and plate tectonics \cite{Conrad2002},\cite{White1989}.Mantle energy is particularly crucial among the Earth's energy sources.A comprehensive understanding of how the Earth generates energy and the current status of its energy sources is indispensable in all disciplines of Earth science.

Seismology has been instrumental in uncovering the physical properties of the Earth, including its layered structure comprising the crust, mantle, and core. Geochemistry has successfully developed models for the Earth's overall chemical composition. Through the integration of physical and chemical data, researchers have achieved the determination of structures and chemical compositions of rocks in each Earth layer.Nonetheless, the chemical composition of the Earth's deepest regions, including the mantle, remains beyond reach, impeding direct observations of the planet's overall chemical composition.

Geoneutrinos, which are antineutrinos emitted from nuclei undergoing $\beta^{-}$ decay of $^{238}$U, $^{232}$Th, and $^{40}$K inside the Earth, offer valuable insights into the concentrations of heat-producing elements throughout the entire planet.The flux of geoneutrinos at any location on Earth is contingent upon the distribution of radioactive elements within the planet.Geoneutrinos have been successfully detected by the 1-kiloton Kamioka Liquid Scintillator Antineutrino Detector (KamLAND) and the 0.3-kiloton Borexino detectors \cite{Kamland2005,Kamland2008,Kamland2013,Borexino2010,Borexino2013,Borexino2015,Borexino2020}, although the limited statistical data has constrained measurements.Situated in Canada, SNO+ will become the third geoneutrino measurement facility in the world\cite{SNO+2016}.The Jiangmen Underground Neutrino Observatory (JUNO) will contribute to the geoneutrino experiment and offer enhanced precision in measurements\cite{JUNO2013,JUNO2016,JUNOgeostrti2015,JUNOgeogigj2019,JUNOgeoJULoc2020}.In its first year of operation, JUNO is expected to record a greater number of geoneutrino events than all preceding detectors combined, enabling the discernment of the relative contributions of \text{U} and \text{Th} with ample statistical data.

%Predicting the geoneutrino signal of JUNO, particularly with high prediction accuracy, is essential for translating particle physics measurements into meaningful geoscientific data. This prediction provides an estimate of the expected geoneutrino signal of JUNO, which is a crucial reference for testing different compositional models of the mantle. Moreover, comparisons of predictions and real data can verify various crustal models, disentangle the mantle signal, and determine the relative contributions of \text{U} and \text{Th}.

Accurate prediction of the geoneutrino signal of JUNO is crucial for bridging particle physics measurements with geoscientific data, ensuring meaningful interpretation.This prediction serves as a vital reference for evaluating various compositional models of the mantle by estimating the geoneutrino signal expected from JUNO.Furthermore, comparing predictions with actual data enables the validation of different crustal models, separation of the mantle signal, and determination of the relative contributions of \text{U} and \text{Th}.

Initial sensitivity studies of JUNO, utilizing the Earth's global composition \cite{JUNO2016,Han:2016geoPotential}, projected an anticipated geoneutrino signal of $39.7^{+6.5}_{-5.2}$ TNU. These projections relied on the "CRUST 2.0" crustal model, the "CUB 2.0" global shear-velocity model, and the high-resolution Moho depth map called "GEMMA" \cite{Bassin2000, Laske2001, Shapiro2002, Reguzzoni2009, Negretti2012}. 
However, these predictions unveiled the insufficiency of global models in terms of accuracy, underscoring the necessity for local high-precision models.

%However, the regional crust within 500 km of JUNO contributes more than 50\% to the total signal\cite{Mantovani2004}, making it necessary to have local refined geological models for a more accurate estimation of the crustal signal. To address this issue, recent studies have proposed the Gravity Field and steady-state Ocean Circulation Explorer (GOCE) Inversion for Geoneutrinos at JUNO (GIGJ) model by Reguzzoni et al.\cite{JUNOgeogigj2019}. This 3-D crustal model was built by inverting GOCE gravimetric data for the $6^{\circ} \times 4^{\circ}$ area centered on the location of the JUNO experiment. Although this model focused on constructing a geophysical and gravimetric model of the local crust, the heat-producing element (HPE) abundances were still based on the homogeneous global model.

Nevertheless, the crust in the JUNO vicinity, spanning a radius of 500 km, accounts for over 50\% of the total signal\cite{Mantovani2004}. Hence, it is imperative to develop localized, refined geological models to enhance the precision of crustal signal estimation.
In response to this challenge, Reguzzoni et al. \cite{JUNOgeogigj2019} introduced the Gravity Field and steady-state Ocean Circulation Explorer (GOCE) Inversion for Geoneutrinos at JUNO (GIGJ) model.The GIGJ model is a three-dimensional crustal model constructed by inverting GOCE gravimetric data within a $6^{\circ} \times 4^{\circ}$ region centered on the JUNO experiment site. While the primary focus of this model was to develop a geophysical representation of the local crust, the abundances of heat-producing elements (HPE) were still derived from a homogeneous global model.

%Gao et al. \cite{JUNOgeoJULoc2020} developed the first refined 3-D JUno LOCal model (JULOC), which is a local crust model of the closest $10^{\circ} \times 10^{\circ}$ grid surrounding JUNO, using local geological, geochemical, and geophysical data. The JULOC model has enabled the construction of a crustal composition model cell-by-cell, reducing the uncertainty in estimating geoneutrino flux in the crust, particularly for the top layer of the upper crust, where the uncertainty has been reduced to 5\%. However, due to limited geophysical and geochemical data, only the top layer of the upper crust has been modeled with the cell-by-cell composition model, while the basement layer has been estimated based on the composition of low-grade metamorphic rocks, resulting in an uncertainty as high as 45\%, with correlated errors between different cells. Furthermore, the boundary between the top layer and the basement layer has been arbitrarily assigned to the uppermost 5 km. To address these issues, an updated integrated JULOC crustal model (JULOC-I) has been developed, incorporating more sample data and a new building method.
Gao et al. \cite{JUNOgeoJULoc2020} introduced the pioneering refined 3-D JUno LOCal model (JULOC), a localized crustal model encompassing the nearest $10^{\circ} \times 10^{\circ}$ grid surrounding JUNO. This model was constructed by incorporating local geological, geochemical, and geophysical data.The JULOC model facilitates a cell-by-cell construction of the crustal composition model, significantly reducing the uncertainty associated with geoneutrino flux estimation within the crust. Notably, the uncertainty in the top layer of the upper crust has been diminished to 5\%.Nonetheless, constrained by the scarcity of geophysical and geochemical data, the cell-by-cell composition model has been exclusively applied to the top layer of the upper crust. For the basement layer, estimation relies on the composition of low-grade metamorphic rocks, leading to an uncertainty as high as 45\%, accompanied by correlated errors among distinct cells. Additionally, the demarcation between the top layer and the basement layer has been arbitrarily set at the uppermost 5 km. To tackle these challenges, an updated integrated JULOC crustal model (JULOC-I) has been formulated, integrating a greater volume of sample data and employing a novel building method.

%This paper presents the JULOC-I model, including its data sources, building method, and the expected geoneutrino signal with associated uncertainty. The expected geoneutrino signals of the four crustal models (Global, JULOC, JULOC-I, and GIGJ) are also summarized, with the potential of JUNO measurements being used to distinguish between these models.

The article presents the JULOC-I model, elaborating on its data sources, building method, and the geoneutrino signal, including associated uncertainty. Additionally, the anticipated geoneutrino signals of the four crustal models (Global, JULOC, JULOC-I, and GIGJ) are summarized, and the potential of JUNO measurements to differentiate between these models is highlighted.

\section{JULOC-I: An Integrated Refined Local 3-D Crustal Model}
\begin{figure} 
   \centering
   \includegraphics[width=0.46\textwidth]{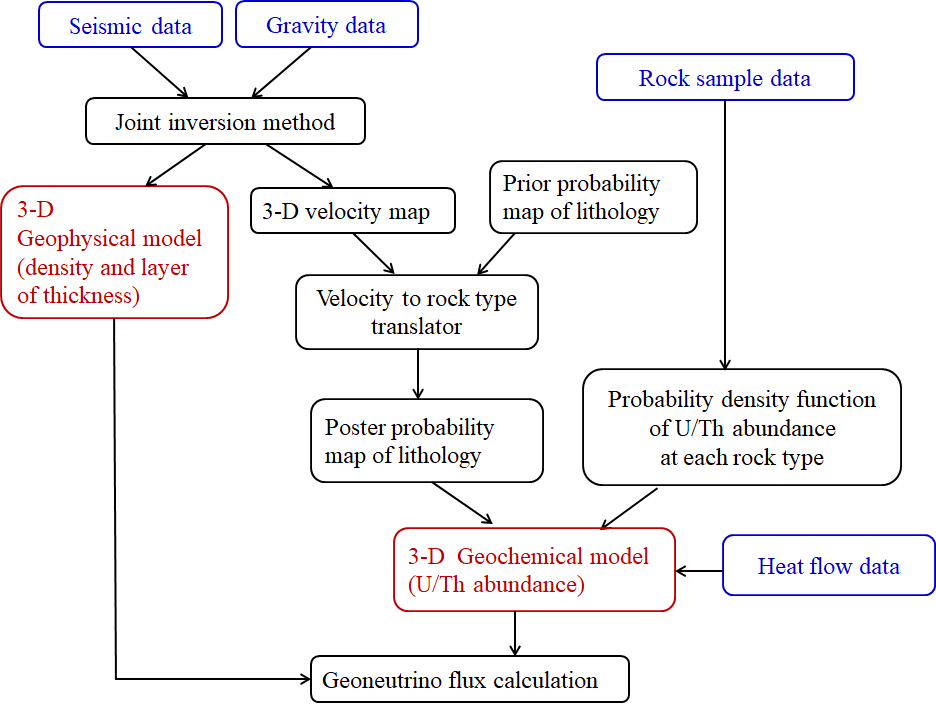} 
   \includegraphics[width=0.46\textwidth]{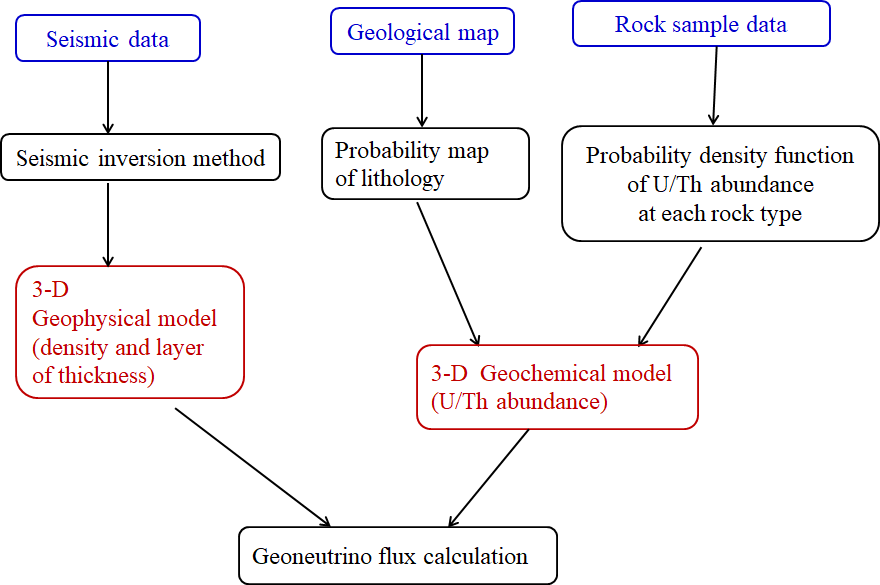}
    \caption{The left plot shows the flowchart that illustrates the construction process of the JULOC-I model. The blue components represent the utilized data, while the red components correspond to the constructed model. The construction process of JULOC is presented on the right for comparison. }
    \label{fig:flowchart}
\end{figure}

This study introduces JULOC-I, an enhanced iteration of the JULOC model, incorporating advancements in data and methodology when compared to its predecessor (refer to Fig. \ref{fig:flowchart}). Notably, several significant distinctions exist between these two models:

%Geophysics model: The JULOC model used 450 permanent broadband seismic stations in South China, while the JULOC-I model used an additional 150 seismic stations. The JULOC model used seismic ambient noise tomography methodology, whereas the JULOC-I model uses joint tomographic inversion methodology with seismic surface wave dispersion and gravity data to obtain more reliable crustal 3-D shear wave structures (S wave).

Geophysical Models: The JULOC model employed 450 permanent broadband seismic stations in South China, while the JULOC-I model expanded this network with an additional 150 seismic stations. In contrast to the JULOC model, which utilized seismic ambient noise tomography methodology, the JULOC-I model utilizes a joint tomographic inversion approach incorporating seismic surface wave dispersion and gravity data. This advanced methodology enhances the reliability of the obtained 3-D shear wave structures (S-waves) within the crust.

Geochemical Model: In addition to utilizing data sourced from literature, such as JULOC, JULOC-I incorporates a sampling survey to address cells with incomplete or missing data regarding rock type. JULOC-I employs a distinct model building method that integrates the crustal 3-D S-wave structures determined in step 1 with a priori probabilities of lithology. This integration facilitates the generation of a posteriori probability lithology map for each cell at 1 km intervals of depth. By considering the concentrations of \text{U} and \text{Th} in each rock type alongside the probability lithology map, a comprehensive 3-D composition probabilistic map is obtained.

Geothermal Data Validation: The probability map and \text{U/Th} abundance obtained from the geochemical model are subsequently constrained by 209 heat flow data points collected around JUNO. This constraint enables the generation of the final probability map for calculating geoneutrino flux. The \text{U}, \text{Th}, and \text{K} data for each layer, in conjunction with their respective heat production and one-dimensional heat transfer function, serve as means to assess the reliability of the geochemical model.

\subsection{Study Area and Sampling Survey}
 \label{subsection:sampling}

\begin{figure}
   \centering
   \includegraphics[width=0.4\textwidth]{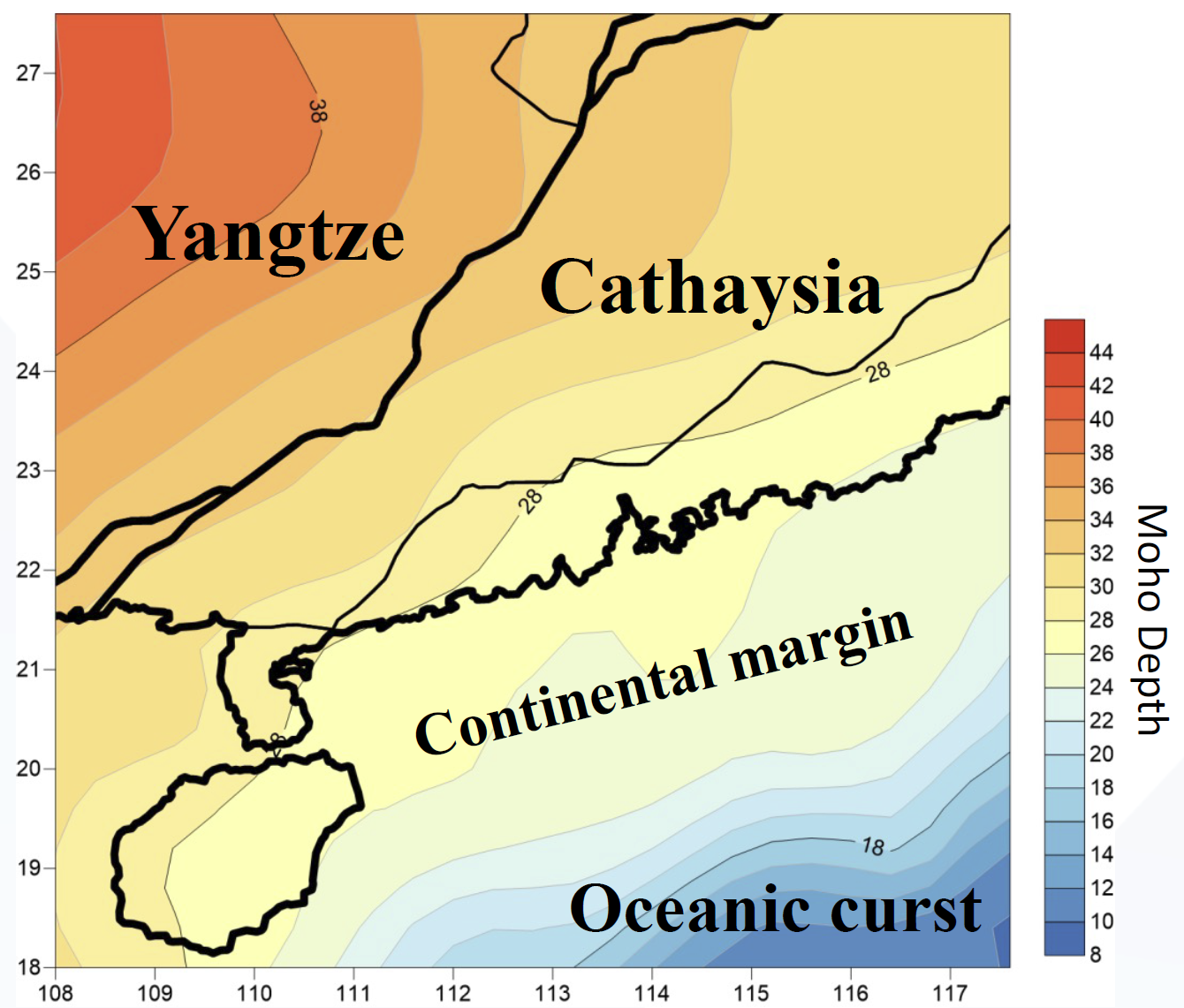} 
   \includegraphics[width=0.45\textwidth]{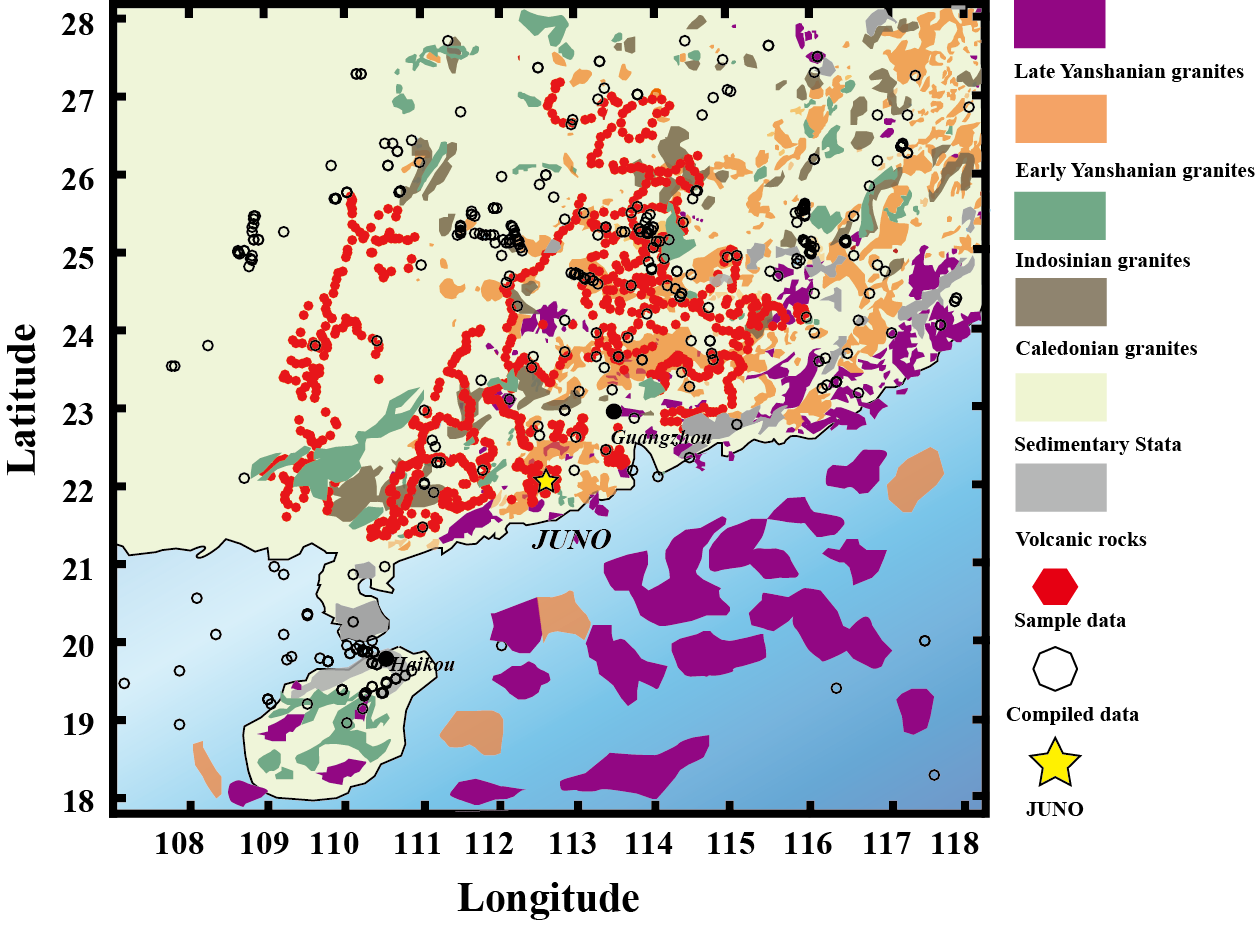} 
    \caption{The left figure displays the four main study areas: Yangtze Block, Cathaysia Block, South China Sea, and the continental margin. The right figure shows the locations where 1002 rock samples were collected within a 500-kilometer radius of JUNO. All the samples were collected from fresh outcrops in the upper crust layer and projected onto a geologic map.}
    \label{fig:samplinglocations}
\end{figure}

 The JUNO detector is located on the coastline of South China ($N 22.12^{\circ}, E 112.52^{\circ}$). Both the JULOC-I and JULOC models focus on a similar research area, which encompasses the $10^{\circ} \times 10^{\circ}$ grid surrounding the detector. Specifically, the area spans from $N 18^{\circ}$ to $N 28^{\circ}$ and from $E 107.2^{\circ}$ to $E 118^{\circ}$. This study area includes the South China Block (Yangtze Block and Cathaysia Block) in the northwest and the South China Sea in the southeast, and the northern continental margin, as left figure in Fig.\ref{fig:samplinglocations}
 
  Due to challenging survey conditions, oceanic data primarily relied on literature sources. However, to ensure adequate data coverage in each cell of the research area, six sampling surveys were conducted, resulting in the collection of over 1,002 rock samples. The accuracy of the JULOC model's cell-by-cell calculation method relies on having a sufficient number of samples per cell to minimize error discrepancies between cells and generate a more precise estimation of geoneutrino flux \cite{TAKEUCHI2019,JUNOgeoJULoc2020}. To accomplish this, the sampling range was divided into $0.4^{\circ}\times 0.4^{\circ}$ grids, with 10 to 20 representative lithology samples collected per cell. However, if adequate data could be sourced from the literature, this step was omitted.

Another survey was conducted to gather lithology samples that were absent in the JULOC model, such as meta-sedimentary and meta-igneous rocks, Precambrian to Quaternary sedimentary rocks with minor volcanic components, and granitic intrusions. For each sample, GPS coordinates and geological data were recorded. Most samples were obtained from representative geological formations at fresh outcrops and securely stored in sealed bags. Right plot in Fig. \ref{fig:samplinglocations} provides an overview of the locations where the 1,002 rock samples were collected.

\subsection{Joint Tomographic Inversion of Seismic and Gravity Data }

\begin{figure}
    \centering
    \includegraphics[width=0.45\textwidth]{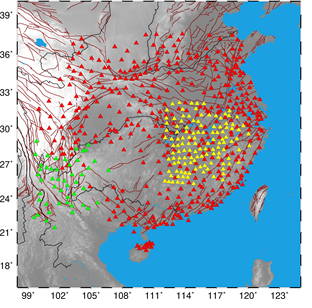}
    \includegraphics[width=0.45\textwidth]{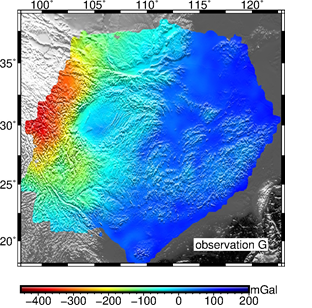}
    \caption{The left plot displays the distribution of seismic stations from which data were collected for the JULOC-I model in South China. The red and green triangles represent permanent seismic stations of the China Earthquake Networks Center, while the yellow triangles represent temporary seismic stations, all of which were used in the JULOC-I model. The right plot shows the distribution of gravity anomalies in the study area, which were obtained from the WGM2012 data set. These gravity anomalies were used as a part of the joint inversion method to obtain more reliable crustal 3-D shear wave structures.}
    \label{fig:sesimicgravity}
\end{figure}

 JULOC was the first 3-D crustal imaging study of South China using one-year continuous seismic ambient noise data collected from 450 permanent broadband seismic stations. The JULOC-I study improved upon this by incorporating an additional 150 seismic stations. JULOC utilized ambient noise tomography, a popular method for imaging the 3-D shear wave structure of the Earth's crust. However, using a single type of geophysical data can result in an incomplete data set due to its sensitivity to only a few physical parameters in a specific region. To overcome this problem, different geophysical data sets can be combined in a joint inversion method. Gravity data is a significant component of such a multidata set joint inversion method, particularly in the context of joint inversion with seismic data. The newly developed JULOC-I model used a joint inversion method of surface wave dispersion and Bouguer gravity anomaly data to obtain more reliable 3-D shear wave structures of the crust \cite{DU2021}. The gravity data in this study came from the WGM2012 global satellite Bouguer gravity anomaly model \cite{Balmino2012}, which is based on the EGM2008 global satellite gravity model. The DTU10 digital elevation model was used to determine $1^{\circ}\times 1^{\circ}$ terrain correction results \cite{Pavlis2012} (see Fig. \ref{fig:sesimicgravity}).

\begin{figure}
    \centering
    \includegraphics[width=0.90\textwidth]{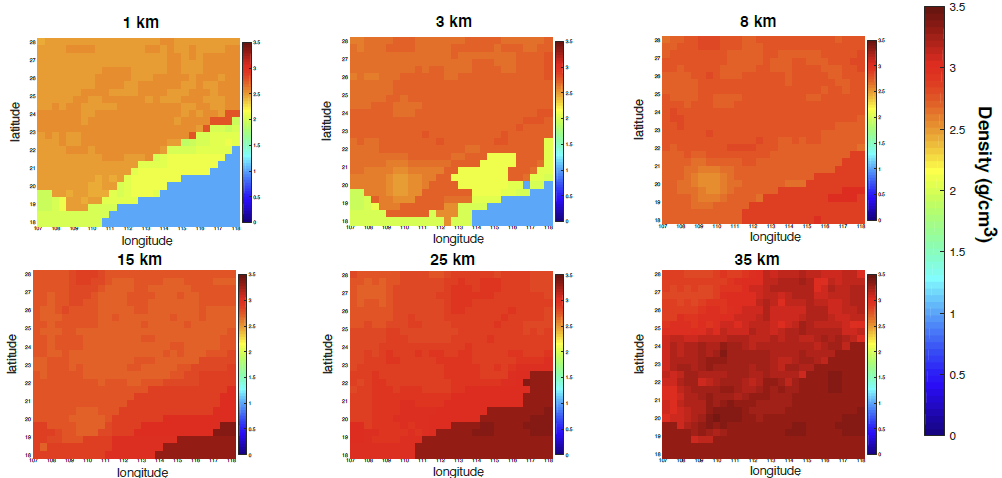}
    \caption{The figure displays the 3-D density structures of the South China Block, which were derived from the 3-D velocity model using a gravity-seismology joint inversion method based on the empirical relationship between the density and velocity of crustal materials at different depth.}
    \label{fig:jointdensity}
\end{figure}

To determine the 3-D density in the South China Block, the same methodology as used in JULOC has been employed, which involves using empirical formulae to transfer S-wave velocity to density \cite{Brocher2005}. Fig. \ref{fig:jointdensity} displays the 3-D density distributions at different depth.
The 3-D model's crustal layering is the same as in the JULOC model, with the upper crust (UC) and middle crust (MC) of the CRUST1.0 model being adopted. However, a different method was used to determine the Moho depth. Specifically, the preferred model was derived from Bouguer gravity inversion under the constraints of approximately 120 deep seismic sounding (DSS) profiles \cite{Hao2014}. This method was chosen because it provides better resolution and coverage of the research area than other available models. The Moho depth is a critical parameter in crustal modeling as it separates the crust from the mantle.

This part only described method and results pertaining to the Yangtze Block and Cathaysia Block. For the South China Sea and the continental margin, the density and layer thickness are determined using the CRUST1.0 values.

\subsection{3-D Map of U/Th Concentrations}
We constructed a 3-D model of \text{U} and \text{Th} abundances by combining the \text{U/Th} abundance of each rock type with the 3-D lithology proportion map based on the geophysical 3-D crustal model. To accomplish this, we used over 500 rock samples to fill in the data gaps in the literature, such as for sedimentary rocks and carbonate rocks, or to rectify uneven data point distribution in cells. The samples were analyzed for trace elements using the Thermo Scientific$^{TM}$ Element XR$^{TM}$ High Resolution ICP-MS system, following the procedure described in \cite{SUNY2022}. 
Table \ref{tab:UThabunYangtze} and Table \ref{tab:UThabunCathia} provide the minimum, median, and maximum\text{U/Th} concentrations in the Yangtze and Cathaysia blocks, respectively, along with the sample sizes. As an example, in the Cathaysia block (Table \ref{tab:UThabunCathia}), the median values of \text{U} (9.28 ppm) and \text{Th} (28.80 ppm) for 685 GRA (Granite-Granodiorite) samples were slightly higher than those reported in the literature (7.49 ppm and 26.0 ppm for \text{U} and \text{Th}, respectively). MGW (siltstone, sandstone, conglomerate, and shale) had low median abundances of 2.64 ppm and 11.00 ppm for \text{U} and \text{Th}, respectively. The observed  \text{U} and \text{Th} abundances in MGW  consistent with values reported in the literature (2.34 ppm and 9.21 ppm, respectively). 

\begin{table}[h!]
\renewcommand\arraystretch{1.5}
\setlength\tabcolsep{14pt}
    \begin{center}
    \caption{Minimum (min), median (median) and maximum (max) \text{U-Th} concentrations in Different Rock Types in the Yangtze Block.}
    \begin{tabular}{| m{2cm}<{\centering}|c|c|c|c|c|}
         \hline
         Rock type & Concentration & Mini & Median & Max & Samples \\
         \hline
        \multirow{2}{*}{GRA} & U(ppm) & 0.02 &3.27 & 36.50  & 833\\
             & Th(ppm) & 0.02 & 13.14 & 84.19 & 851\\
         \hline
         \multirow{2}{*}{MGW} & U(ppm) &0.50 & 	2.20 & 	15.50 &	118\\
            & Th(ppm) & 2.80 &	10.70 &	31.00 &	126 \\
         \hline
         \multirow{2}{*}{AMP} & U(ppm) & 0.08 & 	0.18 & 	0.52 & \multirow{2}{*}{17}\\
             & Th(ppm) & 0.39 & 0.88 &	3.20 & \\
         \hline
         \multirow{2}{*}{BAS} & U(ppm) & 0.02 & 	0.95 & 	13.20  & \multirow{2}{*}{608} \\
          & Th(ppm) & 0.06 & 	4.00  & 	38.30 & \\
         \hline
         \multirow{2}{*}{BGN-GGN} & U(ppm) & 0.02 &	0.82 &	2.68 & \multirow{2}{*}{52}\\
             & Th(ppm) &  0.02 &	7.32 &	37.50 & \\
         \hline
         \multirow{2}{*}{GAB} & U(ppm) & 0.02 &	0.68 &	55.60 & \multirow{2}{*}{110}\\
             & Th(ppm) &  0.05 &	3.53 &	117.00 & \\
         \hline
         \multirow{2}{*}{MGR} & U(ppm) & 0.01 & 0.04 & 0.36 & \multirow{2}{*}{73}\\
             & Th(ppm) &  0.04 &  0.15  & 0.63 & \\
         \hline
         \multirow{2}{*}{MTL} & U(ppm) & 0.01 & 0.08 & 4.24 & 114\\
             & Th(ppm) &  0.01 &  0.10  & 19.60 & 167 \\
         \hline
         \multirow{2}{*}{SLT-PHY-QSC} & U(ppm) & 0.06 & 1.98 & 5.33 & \multirow{2}{*}{40}\\
             & Th(ppm) &  0.11 &  10.93  & 27.50 &\\
         \hline
         
    \end{tabular}
    \label{tab:UThabunYangtze}
     \end{center}
\end{table}

\begin{table}[h!]
\renewcommand\arraystretch{1.5}
\setlength\tabcolsep{14pt}
    \begin{center}
    \caption{Minimum (min), median (median) and maximum (max) \text{U-Th}  concentrations in different rock types in the Cathaysia Block.}
    \begin{tabular}{| m{2cm}<{\centering}|c|c|c|c|c|}
         \hline
         Rock type & Concentration & Mini & Median & Max & Samples \\
         \hline
        \multirow{2}{*}{GRA} & U(ppm) & 0.56 & 9.28 & 73.50  & 685\\
             & Th(ppm) & 0.99& 28.80 & 116.00 & 693\\
         \hline
         \multirow{2}{*}{MGW} & U(ppm) &0.07 & 2.64 & 267.00 & 467\\
             & Th(ppm) & 0.10 & 11.00  & 120.49 & 455 \\
         \hline
         \multirow{2}{*}{AMP} & U(ppm) & 0.07 & 0.41  & 1.99 & \multirow{2}{*}{42}\\
             & Th(ppm) & 0.34 & 1.57  & 9.23 & \\
         \hline
         \multirow{2}{*}{BAS} & U(ppm) & 0.02  & 0.86  & 8.80 & 588 \\
          & Th(ppm) & 0.04 & 3.45  & 30.90 & 603 \\
         \hline
         \multirow{2}{*}{BGN-GGN} & U(ppm) & 1.00 & 3.00 & 8.00 & 21\\
             & Th(ppm) &  2.00 &  13.54  & 49.00 & 22 \\
         \hline
         \multirow{2}{*}{GAB} & U(ppm) & 0.02 & 0.65 & 3.05 & \multirow{2}{*}{45}\\
             & Th(ppm) &  0.02 &  3.23  & 16.57 & \\
         \hline
         \multirow{2}{*}{MGR} & U(ppm) & 0.01 & 0.05 & 0.36 & \multirow{2}{*}{46}\\
             & Th(ppm) &  0.04 &  0.15  & 0.63 & \\
         \hline
         \multirow{2}{*}{MTL} & U(ppm) & 0.01 & 0.89 & 14.77 & 33\\
             & Th(ppm) &  0.03 &  1.98  & 75.75 & 52 \\
         \hline
         \multirow{2}{*}{SLT-PHY-QSC} & U(ppm) & 0.47 & 3.43 & 33.74 & \multirow{2}{*}{50}\\
             & Th(ppm) &  0.68 &  15.23  & 41.60 &\\
         \hline
         
    \end{tabular}
    \label{tab:UThabunCathia}
     \end{center}
\end{table}

\begin{figure}
%\noindent\includegraphics[width=0.45\textwidth]{U_Ave.png}
%\noindent\includegraphics[width=0.45\textwidth]{Th_Ave.png}
\noindent\includegraphics[width=1.0\textwidth]{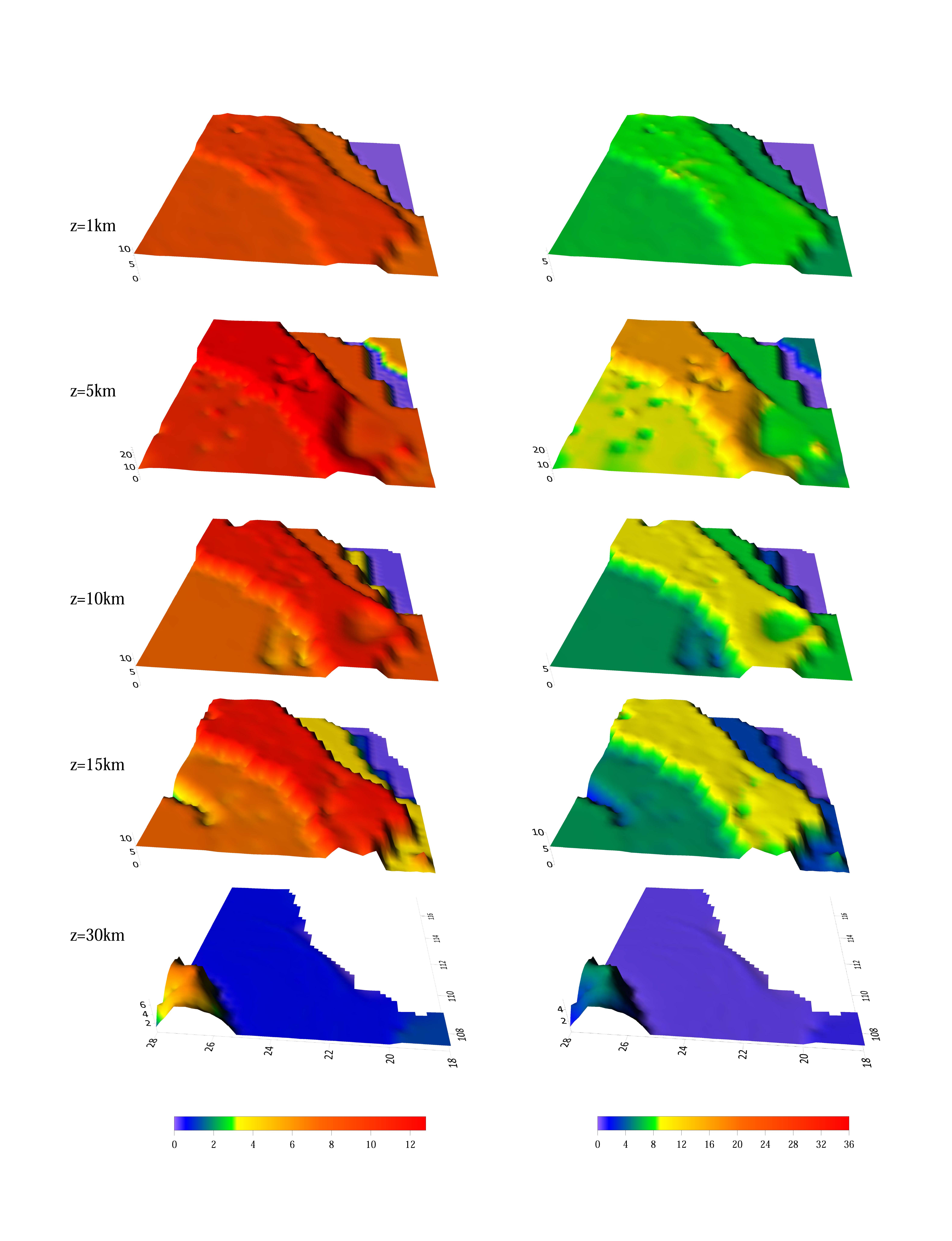}
\caption{The figure displays the abundance of \text{U} and \text{Th} (ppm) in each $0.4^{\circ}\times0.4^{\circ}$ cell of the research area at depths of 1 km, 5 km, 10 km, 15 km, and 30 km, respectively. The data was derived by constructing 3-D probability density functions of \text{U} and \text{Th} concentrations using the method outlined in Appendix B.}
\label{fig:averageabun}
\end{figure}

The method of constructing a 3-D lithology proportion map using 3-D S-wave velocity has previously been utilized by Takeuchi et al. (2019) \cite{TAKEUCHI2019}. This method employs an a priori probability map of lithology, denoted as $P^{x}_{(i)}$, where 'i' denotes the rock type and 'x' denotes the location of the rock, allowing for the consideration of local lithological features. An a posteriori probability lithology map is then generated by establishing a mathematical relationship between the crustal P-wave velocity $V_P^{obs}$ and the proportion of each rock type, as outlined in Appendix A. Using the procedure described in Appendix B, 3-D probability density functions (PDFs) of \text{U} and \text{Th} concentrations are constructed. Figure \ref{fig:averageabun} presents the average \text{U} and \text{Th} abundance in each cell of the research area at depths of 1 km, 5 km, 10 km, 15 km, and 20 km, respectively. Since no sample data were collected for the continental margin area, the \text{U} and \text{Th} abundances were estimated using the empirical relationship between heat production (A) and \text{U/Th} abundances ($[U] \sim 1.5 A$, $[Th] \sim 6A$\cite{Jaupart2014}). Previous geothermal studies predicted different heat production for each layer\cite{LONGZULIE2020,SHAN2011}. The averages and standard deviations of these values were calculated to determine the average \text{U} and \text{Th} abundances along with their corresponding errors.The continental margin area is divided into four depth layers, namely sedimental (U: $2.38\pm0.15 ppm$, Th: $9.50\pm0.59 ppm$), upper crust (U: $2.99\pm0.62$ ppm, Th: $11.96\pm2.49$ ppm), middle crust (U: $1.25\pm0.5$ ppm, Th: $4.98\pm2.00$ ppm), and lower crust (U: $0.42\pm0.15$ ppm, Th: $1.67\pm0.59$ ppm). These values represent the average abundances for each layer.

%, the abundance, density and thickness information are as input value to calculate geoneutrino flux in the next section. 

\subsection{Surface Heat Flow Data to Verify the 3-D \text{U/Th} Concentration Model}
Terrestrial heat flow is a crucial parameter in geophysics as it reflects the thermal state and processes of the Earth's interior. Analyzing the distribution of terrestrial heat flow can offer valuable insights into the geology, geophysics, and geodynamics of the Earth. The two primary sources of terrestrial heat flow are Moho flux, which is the heat from the deep mantle transmitted to the surface, and crustal heat flow, which is the heat generated by the decay of radioactive elements in the crust.

\begin{figure}
    \centering
    \includegraphics[width=0.73\textwidth]{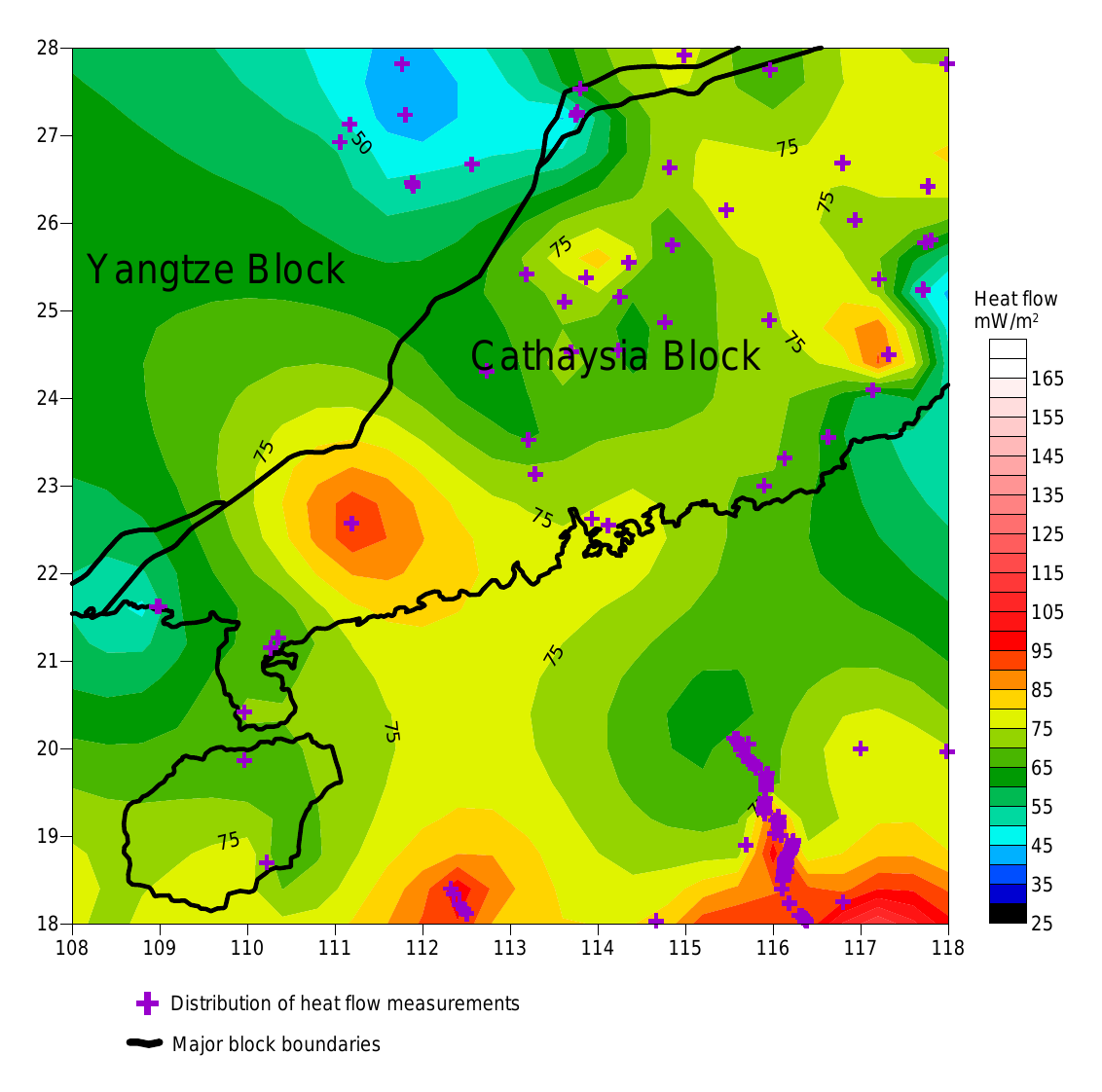}
    \includegraphics[width=0.73\textwidth]{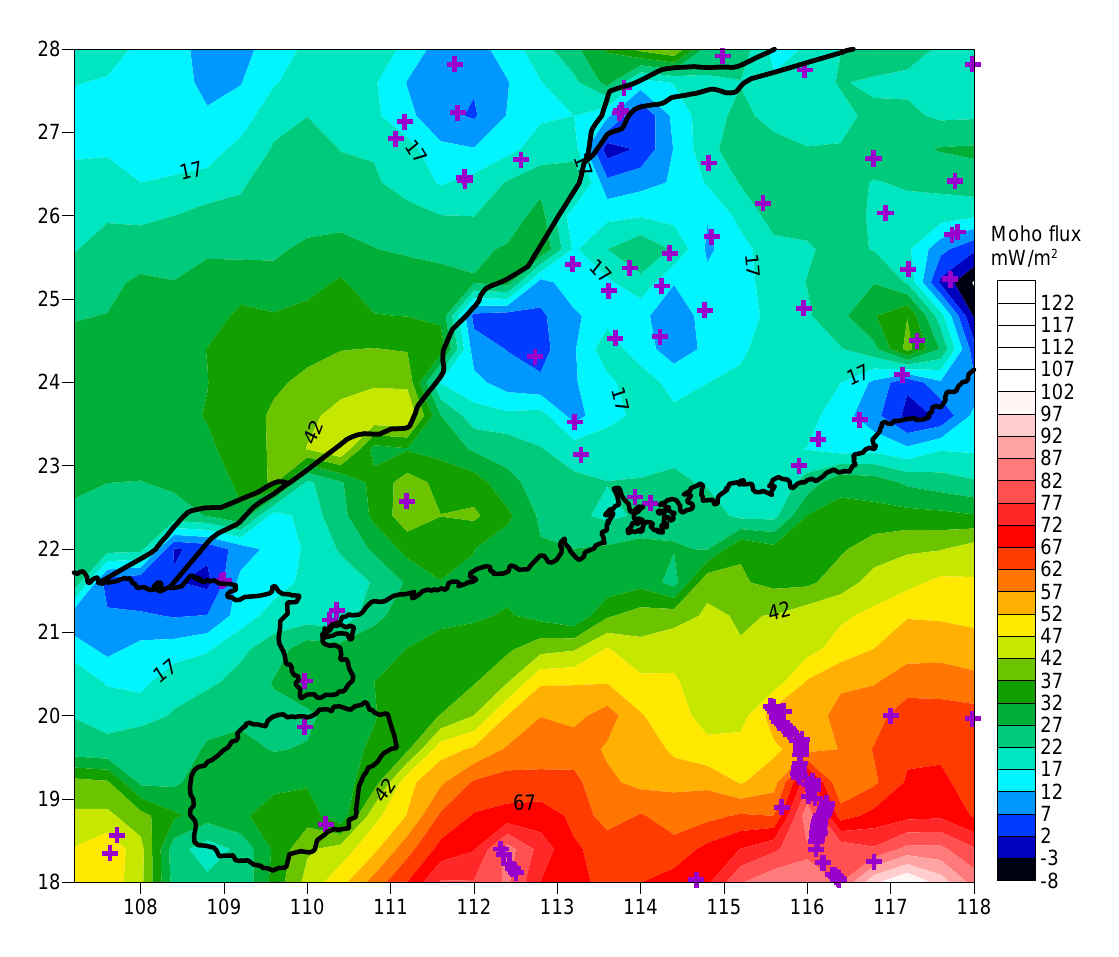}
    \caption{The upper figure illustrates the measured surface heat flow, from which we subtracted the calculated crustal heat flow value based on our established geochemical model to obtain the Moho heat flux, as shown in the lower figure. Despite the study area's high and variable surface heat flux, the Moho heat flux is considered acceptable, which supports the accuracy of our geochemical model.}
    \label{fig:heatflow}
\end{figure}

To validate the 3-D \text{U/Th} crustal concentrations model's accuracy and produce the final geoneutrino flux prediction, we calculated the crustal heat flow in the study area using the \text{U}, \text{Th}, and \text{K} data for each layer in the geochemical model, along with its heat production and one-dimensional heat transfer function. We subtracted the calculated crustal heat flow values from the 206 measured heat flow data points (see Figure \ref{fig:heatflow}) to obtain the Moho heat flow map. The Moho heat flow values' reasonableness can be used to validate the 3-D \text{U/Th} concentrations model's accuracy and ensure the final results' reliability. This analysis provided valuable insights into the region's geology and geophysics and assisted in generating the final geoneutrino flux prediction.

The heat generation of radioactive elements in the crust is expressed by the heat generation rate,A, which depends on the abundance of radioactive elements and the density of rock(\cite{Rybach1976}).
\begin{equation}
   A=10^{-5}\rho(9.52C_U+2.56C_{Th}+3.48C_{k});
    \label{Eq:Aheatflow}
\end{equation}

Where A is in $\mu W/m^3$, $\rho$ is the density in $kg/m^3$. $C_U,C_{Th}$ and $C_K$ are the abundances of U(ppm), Th (ppm) and K(\%), respectively. 

The conduction direction of heat flow is from high temperature region to low temperature region. In general, the temperature inside the Earth is higher than the surface, so the direction of heat transfer from the Earth is almost perpendicular to the surface. In one-dimensional steady-state conditions, the heat flow at one specific depth is always equal to the surface heat flow $q_0$ minus the total heat production above that depth z (\cite{Lachenbruch1970}) 
\begin{equation}
q(z)= q_{0}-\int_0^z A(z).
\end{equation}
where A(z) is heat production at the depth z. 
Specifically, the Moho flux q(M) is the heat flow at the Moho (z = H).
\begin{equation}
q(M)= q_{0}-\int_0^H A(z). 
\end{equation}
H is the crustal thickness.
Scholars established different global models regarding the ratio of Moho flux to crustal heat flow. 

These global models suggest that the range of Moho flux falls between 11-17 $mW/m^2$ for stable regions (\cite{Rudnick2003,Hacker2011}). In our study, the Moho flux of the study area is shown in Figure \ref{fig:heatflow}, with a value of around 20 $mW/m^2$ on the continent and over 45 $mW/m^2$ for the Oceanic crust. Despite the high and variable surface heat flux in the study area, the Moho flux is considered acceptable. The density model from geophysics and abundance model from geochemistry are both reasonable.

\section{Prediction of Geoneutrino Signal at JUNO by JULOC-I}

The joint geophysical and geochemical 3-D local crustal models developed around JUNO can calculate the Earth's crust's geoneutrino flux. However, predicting the geoneutrino signal detectable at the JUNO detector requires knowledge of the particle's detection via inverse beta decay (IBD) reaction. The mathematical details of geoneutrino signal calculation are outlined in R. Gao et al. (2020) \cite{JUNOgeoJULoc2020} and the parameter settings and error estimation methods are described in Appendix C. Table \ref{tab:Loccontribution} summarizes the expected geoneutrino signals and uncertainties (1 standard deviation) in TNU for JUNO from the local crust (with a $10^\circ \times 10^\circ$ area as in the JULOC-I research area). These results reveal insights into the local crust's contribution to the geoneutrino signal detected by JUNO, which is crucial for comprehending the Earth's geology and geophysics, especially for extracting information about the mantle.

\begin{table}[h!]
\renewcommand\arraystretch{1.5}
\setlength\tabcolsep{18pt}
    \begin{center}
    \caption{The predicted geoneutrino signal (S) from the local crust (LOC) in the JUNO research area and the uncertainty ($\sigma$) in TNU are presented, including contributions from the continental crust (CC: upper crust (UC), middle crust (MC), and lower crust (LC)), continental margin(CM) and oceanic crust (OC).}
    \begin{tabular}{|c|c|c|c|}
         \hline
          & $S_{\text{U}} \pm \sigma$ & $S_{\text{Th}}  \pm \sigma$ &  $S_{\text{U+Th}} \pm \sigma$ \\
         \hline
        UC & $19.48^{+2.31}_{-1.94}$  & $4.25^{+0.25}_{-0.26}$ & $23.81^{+2.39}_{-1.93}$ \\
         MC & $1.73^{+1.56}_{-0.97}$  & $0.97^{+1.35}_{-0.67}$ & $2.99^{+1.98}_{-1.40}$ \\
         LC & $0.11^{+0.10}_{-0.08}$ & $0.01^{+0.01}_{-0.01}$ & $0.11^{+0.10}_{-0.08}$\\
        CC  & $21.75^{+2.92}_{-2.27}$ & $5.26^{+1.33}_{-0.69}$ & $27.28^{+3.10}_{-2.52}$ \\
        CM & $1.24 \pm 0.01 $ & $0.44 \pm 0.003$ & $1.67 \pm 0.01$ \\
        OC & $0.2 \pm 0.05$ & $0.1 \pm 0.01$ & $0.3\pm 0.05$ \\
         Total Crust  & $23.19^{+2.92}_{-2.39}$ & $5.80^{+1.33}_{-0.69}$ & $29.25^{+3.10}_{-2.52}$ \\
    \end{tabular}
    \label{tab:Loccontribution}
     \end{center}
\end{table}

The expected total geoneutrino signal at JUNO is the sum of four independent components: signals from the bulk crust, the continental lithospheric mantle (CLM), the depleted mantle (DM), and the enriched mantle (EM). The bulk crustal signal comprises the signal from two parts of the local crust (LOC) and the far-field crust (FFC). Table \ref{tab:totcontri} presents the predicted total geoneutrino signal at JUNO. These findings provide essential insights into the contributions of distinct Earth layers to the geoneutrino signal detected by JUNO.

\begin{table}[h!]
    \renewcommand\arraystretch{1.5}
\setlength\tabcolsep{20pt}
    \begin{center}
    \caption{The table summarizes the expected geoneutrino signals and uncertainties (1 standard deviation) in TNU for JUNO, including those from uranium ($S_\text{U}$ ), thorium ($S_\text{Th}$), and their sum ($S_{TOT}$). The bulk crust contribution consists of contributions from the local crust (LOC) and far-field crust (FFC). The lithosphere contribution is the sum of the bulk crust and continental lithospheric mantle (CLM) contributions. The total signal is the sum of the lithosphere, depleted mantle (DM), and enriched mantle (EM) contributions. All data originate from the JULOC model except for the LOC contribution.}
    \begin{tabular}{|c|c|c|c|}
        \hline
           & $S_{\text{U}} \pm \sigma$ & $S_{\text{Th}}  \pm \sigma$ &  $S_{\text{U+Th}}  \pm \sigma$ \\
           \hline
          %LOC & 24.7 \pm 3.3 & 5.8 \pm 0.4 & 30.7 \pm 3.2\\
          LOC   & $23.19^{+2.92}_{-2.39}$ & $5.80^{+1.33}_{-0.69}$ & $29.25^{+3.10}_{-2.52}$ \\
          FFC  & $7.6 \pm 1.6$ & $2.3 \pm 0.2$ & $9.8 \pm 1.7$\\
         % Bulk crust & 32.3 \pm 3.7 & 8.1 \pm 0.5 & 40.5 \pm 3.6 \\
          Bulk crust & $30.79^{+3.33}_{-2.88}$ & $8.10^{+1.34}_{-0.72}$ & $39.05^{+3.54}_{-3.04}$\\
          CLM  & $1.3 ^{+2.4}_{-0.9}$  & $0.4 ^{+1.0}_{-0.3}$ & $2.1 ^{+3.0}_{-1.3}$ \\
         % Lithosphere & 33.6 ^{+4.4}_{-3.8} & 8.8^{+1.1}_{-0.6} & 42.6^{+4.7}_{-3.8} \\
          Lithosphere & $32.09^{+4.10}_{-3.02}$ & $8.5^{+1.67}_{-0.78}$ & $41.2^{+4.64}_{-3.31}$ \\
          DM & $4.2$ & $0.8$ & $4.9$ \\
          EM & $2.9$ & $0.9$ & $3.8$ \\
         Total & $39.19^{+4.10}_{-3.02}$ & $10.2^{+1.67}_{-0.78}$ & $49.9^{+4.64}_{-3.31}$
    \end{tabular}
    
    \label{tab:totcontri}
    \end{center}
\end{table}

Table \ref{tab:totcontri} reveals that the local crust (LOC) contributes around 74.9\% of the total crustal geoneutrino signal, with 81.4\% originating from the upper crust (UC) layer while the middle crust (MC) layer contributed 10.2\%. Hence, improving model precision necessitates reducing uncertainty in the UC layers. The uncertainty in the estimated geoneutrino signal primarily stems from inputs in the crustal model, including crustal layer thickness, density, and uranium-thorium (U-Th) abundances. The geoneutrino detection cross-section and electron antineutrino survival probability are expected to induce an uncertainty of ~1\% \cite{MAO2019}, excluded from the calculation.

The uncertainty related to the thickness and density of crustal layers is negligible, as any variation from different cells can be eliminated during calculation, particularly when seismic ambient noise tomography is used to determine the crustal layer thickness and density \cite{JUNOgeoJULoc2020}. Hence, only the uncertainty attributed to \text{U} and \text{Th} abundances was considered, and the methodology used to calculate the geoneutrino uncertainty from abundances was the same as that used for the JULOC model.

If each cell of the UC layer had enough rock samples to effectively describe the abundance distribution, the uncertainty would be uncorrelated. However, with respect to the MC and the lower crust (LC) layers, there were not enough samples, so the uncertainty is considered to be correlated, resulting in the UC having a smaller uncertainty than the MC and LC. Improving the precision of the abundance measurements in the MC and LC layers,especially MC, could help reduce the overall uncertainty in the geoneutrino signal prediction.

\section{Model Comparison and Conclusions}
%With a detector at least 20 times larger than existing detectors, JUNO is poised to join the family of geoneutrino experiments, offering an even better opportunity to measure geoneutrinos. The number of geoneutrinos represent different Earth composition models, thus signal prediction using different geophysical and geochemical models before JUNO comes online is both necessary and meaningful. Moreover, to reduce the uncertainty of geoneutrino flux prediction, the local refined geological models are important, especially as these models also provide the possibility of extracting the mantle component.
%// keep or not?
There are two existing local refined models for the JUNO region: GIGJ (GOCE Inversion for Geoneutrinos at JUNO) \cite{JUNOgeogigj2019} and JULOC, with GIGJ being a refined geophysical model and JULOC being the first comprehensive 3-D high-resolution geophysical and geochemical model. JULOC-I, the third refined model, is an advanced integrated model that incorporates seismic ambient noise, gravity, rock sample, and heat flow data, all of which are constrained to each other. 

JULOC-I estimated the total lithosphere geoneutrino signal to be $41.2^{+4.64}_{-3.31}$ TNU, while the JULOC model predicted an expected signal of $40.4^{+5.6}_{-5.0}$ TNU. The updated model, which includes additional data points, decreased the total expected geoneutrino signal error from 14\% to 11\%. This represents a reduction of 3\%. The expected geoneutrino signal is significantly higher, compared to the expected geoneutrino signal of $30.9^{+6.5}_{-5.2}$ TNU\cite{JUNOgeostrti2015}, based on the global crustal model, and 29.9 TNU from the GIGJ model.

\begin{figure}
    \centering
    \noindent\includegraphics[width=0.45\textwidth]{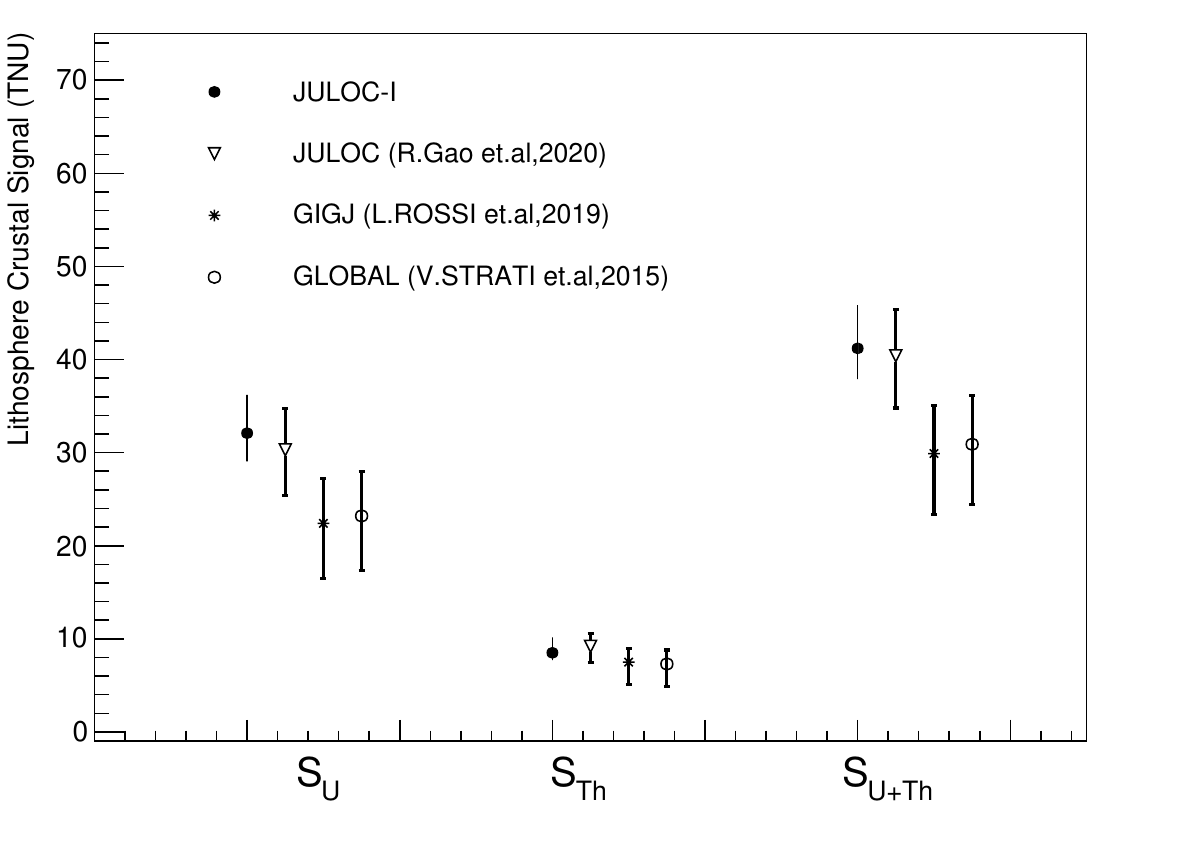}
    \noindent\includegraphics[width=0.45\textwidth]{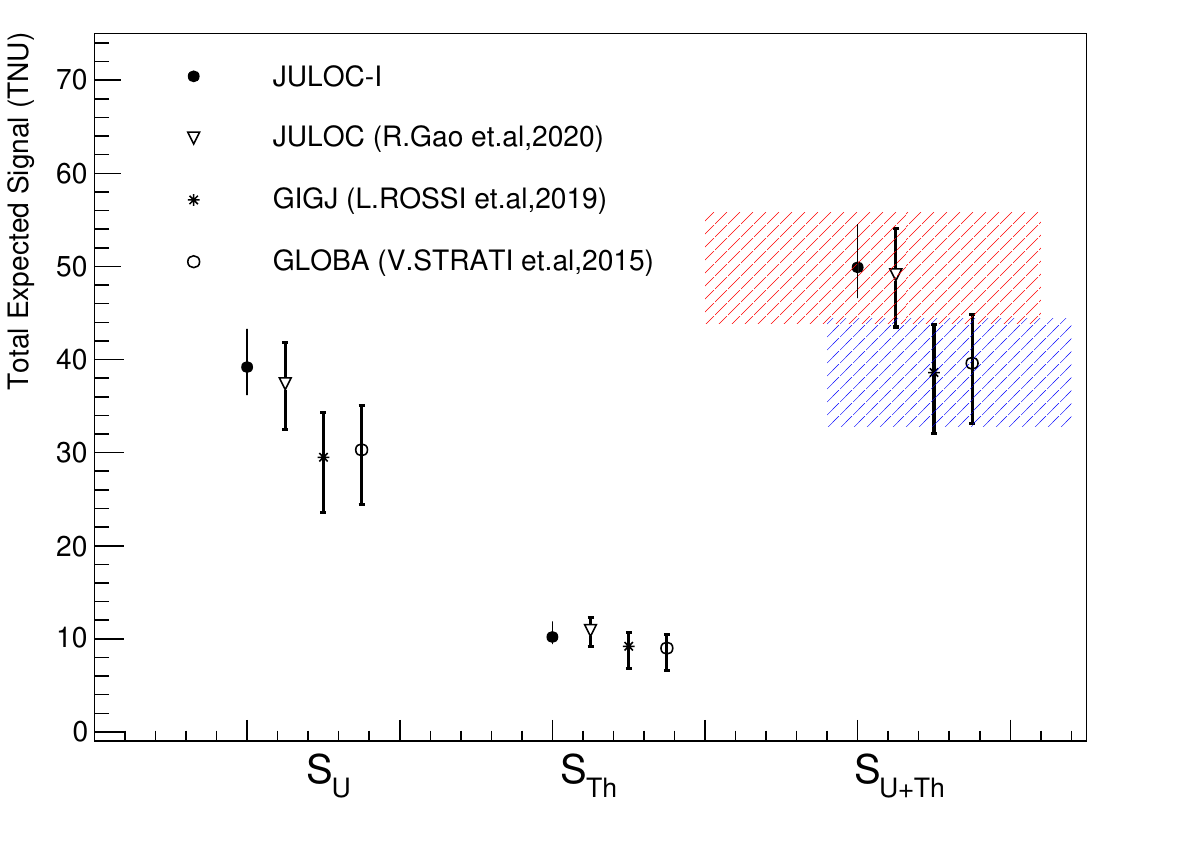}
       \caption{The left plot displays the expected lithosphere geoneutrino signals, $S_{\text{U}}$, $S_{\text{Th}}$, and $S_{\text{U+Th}}$, for different models, from left to right: JULOC-I, JULOC, GIGJ, and Global models (Crust2.0). The error bars indicate the error attributed mainly to \text{U} and \text{Th} abundance. The right plot shows the total expected geoneutrino signal, with the mantle contribution added to the left plot. The shaded areas represent the precision of geoneutrino measurements at JUNO over one year. The mean value of the red-shaded area is based on the JULOC-I model plus mantle signal, while the blue-shaded area is based on the Global model plus mantle signal. The experimental precision is 13\% \cite{Han:2016geoPotential}.}
    \label{fig:comparecrust}
\end{figure}

Figure \ref{fig:comparecrust}'s left-hand plot displays the expected lithosphere geoneutrino signals from the Global (Crust2.0), GIGJ, JULOC, and JULOC-I models. The error bars represent the error mainly attributable to the abundance of \text{U} and \text{Th}. The plot illustrates that the JULOC-I and JULOC models have higher expected geoneutrino signals than the GIGJ and Global models based on the different local composition data. The significant increase in signals is primarily due to the upper crust of the South China Block, which is richer in \text{U/Th} than the global average. This finding aligns with the widespread distribution of high \text{U-Th} granite intrusions in the JUNO area\cite{WANG2001121}. This result is also supported by sampling survey findings and heat flow data.

%The plot also shows the higher expected geoneutrino signals of the JULOC and JULOC-I model are mainly attributable to the \text{U} contribution, with the local $\text{Th}$ contribution being similar to the global average. This may denote that the south of China seems has a higher $\text{U/Th}$ ratio than global average. 

The precision level of the geoneutrino experiment measurements at JUNO is shown in the right-hand plot of Fig. \ref{fig:comparecrust}, along with the total expected signal. The total expected signal is the sum of the lithosphere crustal signal and the mantle signal, with the mantle contribution taken from the published paper on expected geoneutrino signals at JUNO, using a geochemical mantle model \cite{McDonough1995}. The red- and blue-shaded areas represent the experimental precision level, with the mean value of the red-shaded area based on the JULOC-I model plus mantle contribution, and the blue-shaded area based on the Global model plus mantle contribution. Both are considered to have 13\% experimental precision for one year of data, as reported by R. Han et al. (2016) \cite{Han:2016geoPotential}, although it should be noted that different expected geoneutrino numbers may cause this value to differ slightly. The expected signal error is the same as the crustal error in the left-hand plot. The right-hand plot indicates that using the same mantle model and based on one year of data, the measurement precision at JUNO is sufficient to distinguish different crustal models.

\begin{figure}
    \centering
    \noindent\includegraphics[width=0.65\textwidth]{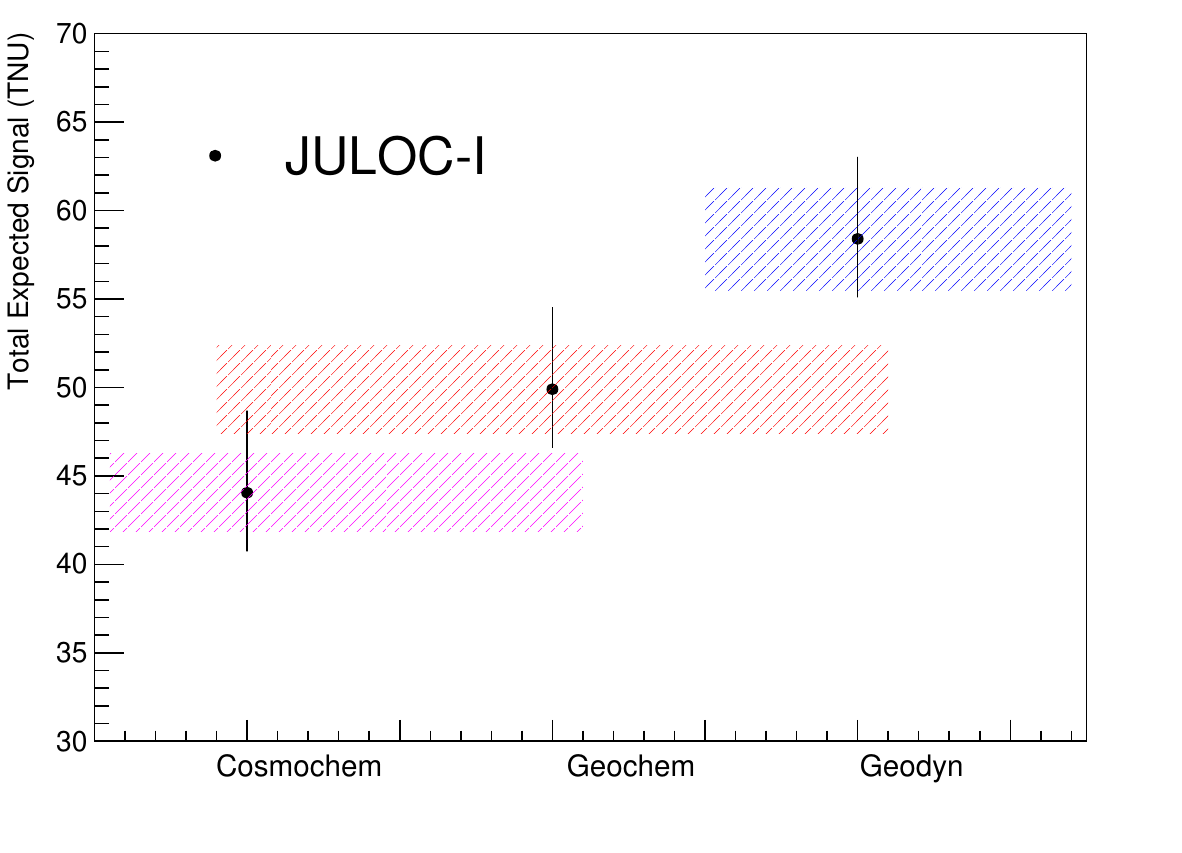}
       \caption{The figure displays the total expected geoneutrino signal in TNU at JUNO based on different mantle models, including the cosmochemical, geochemical, and geodynamical models. The lithosphere contribution to the total expected signal has been estimated for each mantle model using the JULOC-I crustal model. The shaded ranges represent the 10-year precision level of geoneutrino measurements at JUNO, with the pink-shaded area representing the cosmochemical model, the red-shaded area representing the geochemical model, and the blue-shaded area representing the geodynamical model. The 10-year experimental precision is 5\% \cite{Han:2016geoPotential}.}
    \label{fig:comparetotal}
\end{figure}

As previously mentioned, the measurement of geoneutrino signals at JUNO consists of a lithosphere contribution and a mantle contribution. The mantle contribution is predicted based on various assumptions regarding the distribution of highly-potassium-enriched (HPE) materials. All published expected geoneutrino measurements at JUNO are based on geochemical assumptions. However, three classes of compositional estimates exist: cosmochemical \cite{Neill2008,Palme2003}, geochemical \cite{Arevalo2009,McDonough1995}, and geodynamical \cite{Turcotte2002}. These estimates provide different abundances of HPEs and are also referred to as Low-Q (low radiogenic power), Middle-Q (middle radiogenic power), and High-Q (high radiogenic power) \cite{Sramek2013}.

Table \ref{tab:totcontri} shows that the mantle contribution is mainly based on the geochemical model. The lithosphere contribution is added to the mantle results of the cosmochemical and geodynamical models (from Table 4 in \v{S}r{\'a}mek et al. (2013) \cite{Sramek2013}). The total expected geoneutrino signals at JUNO using different mantle models are shown in Fig. \ref{fig:comparetotal}, with the crustal model being JULOC-I. This paper did not include uncertainty in the mantle assumptions, and the error bar was derived only from the lithosphere parts as outlined in Table \ref{tab:totcontri}. The 10-year precision level of geoneutrino measurement at JUNO is plotted as shaded areas in Fig \ref{fig:comparetotal}. The pink-shaded area represents the mean value of the cosmochemical mantle model, the red-shaded area represents the geochemical mantle model, and the blue-shaded area represents the geodynamical mantle model. The 10-year experimental precision is 5\% \cite{Han:2016geoPotential}.

Fig. \ref{fig:comparetotal} shows that different mantle models produce distinct expected geoneutrinos at JUNO. Therefore, sufficient experimental data is necessary to distinguish between the different mantle models and to improve the accuracy of crustal models. With 10 years of data, JUNO will collect enough data to improve the precision level of measurement to 5\%. Currently, the JULOC-I model provides the lowest error at 11\%. Both of these aspects offer JUNO the opportunity to differentiate between different mantle models. Looking into the future, a higher accuracy crustal model will still be necessary not only to test different mantle models but also to constrain the \text{U/Th} ratio.

In conclusion, the JUNO detector offers an exciting opportunity to obtain a statistically significant and precise measurement of geoneutrinos. An accurate estimation of the crustal contribution is an important and necessary step to translate particle physics measurements into geoscientific goals. The first precise, integrated local crustal model based on seismic, gravity, abundance, and heat flow data has been developed, and it predicts geoneutrino flux with the highest accuracy among the four existing crustal models. Therefore, within its first year of operation, JUNO should distinguish the four different local crustal models. Within 10 years of data, precise local crustal models should enable the extraction of the mantle component, which is a highly sought-after goal in geoscience.

\newpage
\section*{List of abbreviations used}
KamLAND: Kamioka Liquid Scintillator Antineutrino Detector \\
JUNO: Jiangmen Underground Neutrino Observatory \\
GOCE: Gravity Field and steady-state Ocean Circulation Explorer \\
GIGJ: GOCE Inversion for Geoneutrinos at JUNO \\
JULOC: JUno LOCal model \\
JULOC-I: JUno LOCal - Integrate model \\
LOC: Local Crust \\
FFC: Far Field Crust \\
CLM: Continental Lithospheric Mantle \\
DM: Depleted Mantle \\
EM: Enriched Mantle \\
UC: Upper Crust \\
MC: Middle Crust \\
LC: Lower Crust \\
CM: Continental Margin\\
OC: Oceanic Crust \\
IBD: Inverse Beta Decay   \\
TNU: Terrestrial Neutrino Units \\
SED: Sediments \\
BAS: Basalt \\
GRA: Granite-Granodiorite \\ 
BGN-GGN: Granite Gneiss, Biotite Gneiss \\
MGW: Siltstone, Sandstone, Conglomerate and Shale \\
SLT-PHY-QSC: Slate, Phyllite, Schist  \\
BGN-GGN: Granite Gneiss, Biotite Gneiss \\
AMP: Amphibolite \\
IGR: Intermediate Granulite \\
MGR: Mafic Granulite \\
GAB: Gabbro \\
MPE: Metapelite  \\
MTL: Mantle Materials \\
\section*{Acknowledgement}
This study is supported by National Natural Science Foundation of China (U1865206,42074092 and 12205018), National Key Research and Development Program of China (2018YFA0404100).

\newpage

\section*{Appendices}

\begin{appendices}

\section{Estimating the lithology proportion by means of 3-D shear-wave velocity }
 \label{subsection:UThdis}
%As mentioned above, JULOC model built a crustal composition model cell by cell to improve the uncertainty of the estimated geoneutrino signal. But this limited by two aspects, one is the number of rock samples in each cell, another is the proportion map of lithology. To get enough rock samples, we collected more data from literature and did the sampling survey to complete the cell which lacking data in literature. In this paper, 500 samples are used from 1002 sampling survey rocks. In combination the freely accessible EarthChem rock database and literature, there are almost 3000 data compiled in the model.
%the data points used to built U and Th model are in Fig.\ref{fig:rawdata}.
%\begin{figure}
%\noindent\includegraphics[width=0.55\textwidth]{Udatapoints.png}
%\noindent\includegraphics[width=0.55\textwidth]{Thdatapoints.png}
%\caption{The data distribution of }
%\label{fig:rawdata}
%\end{figure}
%We then built a 3-D composition crustal model combining the lithology proportion estimation via p-wave velocity data and Bayesian inference with probability density functions (PDFs) of element concentrations of each rock type. 

Estimating the proportion of different lithologies in the subsurface is an important task in geology and geophysics. One method for doing this is by using 3-D shear-wave velocity (Vs) models, which can be obtained through seismic surveys or other geophysical techniques. Different lithologies have different Vs values, so by analyzing the Vs model, it is possible to estimate the proportion of different lithologies in the subsurface. This is because the Vs values are related to the elastic properties of the rocks, which in turn are related to their composition and structure. For example, sedimentary rocks typically have lower Vs values than igneous or metamorphic rocks.

The use of 3-D S-wave velocity for lithology proportion estimation has been previously demonstrated by Takeuchi et al. (2019) \cite{TAKEUCHI2019} and Sun et al. (2022) \cite{SUNY2022}. This method involves assuming a priori probability map of lithology as $P^{x}_{(i)}$, where '$i$' represents the type of rock and '$x$' represents the location of that rock. This prior probability allows local lithological features to be considered.

In the prior probability map, the lithology of the upper crust (UC) is estimated from the geological map, while the lithology of the middle crust (MC) and lower crust (LC) are estimated based on the average crustal composition of East China \cite{Gao1998}. Table \ref{tab:priorprob} shows the a priori probability map, with SED (Sediments), BAS (Basalt), GRA (Granite), BGN-GGN (Granite Gneiss, Biotite Gneiss), MGW (Metagraywacke), and SLT-PHY-QSC (Slate, Phyllite, Schist) in the UC layer. In the MC layer, there are BGN-GGN (Granite Gneiss, Biotite Gneiss) and AMP (Amphibolite). In the LC layer, there are IGR (Intermediate Granulite), MGR (Mafic Granulite), and MPE (Metapelite).

\begin{table}[h!]
\renewcommand\arraystretch{1.5}
\setlength\tabcolsep{18pt}
    \begin{center}
    \caption{The input a priori probability of continental crust rock composition in the research area.}
    \begin{tabular}{|c|c|c|}
         \hline
         UC & MC  & LC \\
         \hline
         SED 0.08  & BGN-GGN 0.1 & IGR 0.6 \\
         BAS 0.02 & AMP 0.9  & MGR 0.2 \\
         BGN-GGN 0.1 &  & MPE 0.2 \\
        GRA 0.4 &  &  \\
        MGW 0.25 &  &  \\
        SLT-PHY-QSC 0.15 &  &   \\
    \end{tabular}
    \label{tab:priorprob}
     \end{center}
\end{table}

An a posteriori probability lithology map is then developed by establishing a mathematical connection between the crustal P-wave velocity $V_P^{obs}$ and rock type proportion. The observed $V_P$ velocity is calculated from the observed S-wave velocity $V_S^{obs}$ using the empirical equation \cite{Brocher2005}, as shown in Eq. \ref{Eq:vstovp}. The S-wave velocity is derived from the 3-D velocity model constructed using the gravity-seismology joint inversion method, as described earlier.
\begin{equation}
\begin{aligned}
    V_P^{obs}(km/s) 
    & = 0.9409 + 2.0947 V_S^{obs} - 0.8206 (V_S^{obs})^2 & \\
    & + 0.2683 (V_S^{obs})^3 -0.0251 (V_S^{obs})^4  &
    \label{Eq:vstovp}
\end{aligned}
\end{equation}
where $V_P^{obs}$ and $V_S^{obs}$ are wave velocities at location 'x' as derived from the geophysical model.
The a posteriori probability of a rock type can be calculated:
\begin{equation}
    P(i|V_P^{obs})=\frac{P(V_P^{obs}|i) P(x|i)}{(\Sigma_{i=1}^n P(V_P^{obs}(x|i) P(x|i)}
    \label{Eq:posteq1}
\end{equation}

where $P(x|i)$ refers to the a priori probability and $P(V_P^{obs}|i)$ represents the probability of $V_P^{obs}$ when a certain rock occurs, $P(V_P^{obs}(x|i))$is sampled from a gaussian distribution with mean and standard deviation summarized in in Tab.4 \cite{Christensen1995,Huang2013}.

\begin{figure}
    \centering
    \noindent\includegraphics[width=0.45\textwidth]{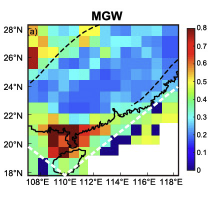}
    \noindent\includegraphics[width=0.45\textwidth]{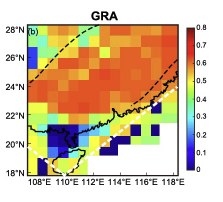}
    
    %\noindent\includegraphics[width=0.3\textwidth]{gra_5km.png}
    %\noindent\includegraphics[width=0.3\textwidth]{bgn_5km.png}
    %\noindent\includegraphics[width=0.30\textwidth]{mgw_5km.png}
    %\noindent\includegraphics[width=0.30\textwidth]{slt_5km.png}
    %\noindent\includegraphics[width=0.30\textwidth]{gab_5km.png}
     %\noindent\includegraphics[width=0.30\textwidth]{mgr_5km.png}
     %\noindent\includegraphics[width=0.30\textwidth]{igr_5km.png}
     %\noindent\includegraphics[width=0.30\textwidth]{mtl_5km.png}
    % \noindent\includegraphics[width=0.30\textwidth]{mpe_5km.png}
     %\noindent\includegraphics[width=0.30\textwidth]{amp_5km.png}
     \caption{Proportions of volume of different 3 km sedimentary rock and 3 km Granite rock type.}
    \label{fig:11530kmpro}
\end{figure}

 \section{Building of \text{U/Th} concentrations cell-by-cell}
Within a 500-kilometer radius of the JUNO center, the region was divided into cells of size $0.4^\circ \times 0.4^\circ \times$1 km. Each cell was further subdivided based on the rock types mentioned in subsection \ref{subsection:sampling}. The \text{U/Th} abundance of each rock type, combined with the 3-D lithology proportion map derived from seismic wave velocity data, was used to calculate the average \text{U/Th} abundance of each cell.

The processor used to calculate the \text{U/Th} abundance of each rock type followed the following criteria: If there were more than five data points available for a rock type in a cell, the data were used to calculate the abundance and related uncertainty. When the available sample size was less than 10, the expected abundance and uncertainty were calculated using the sample mean and standard deviation, respectively. For sample sizes exceeding 10, the expected abundance and uncertainty were determined by fitting a normal, logarithmic, or gamma distribution to the data.
By applying these criteria, the processor was able to accurately estimate the \text{U/Th} abundance of each rock type in each cell, even when the sample size was relatively small.

\begin{figure}
\includegraphics[width=0.9\textwidth]{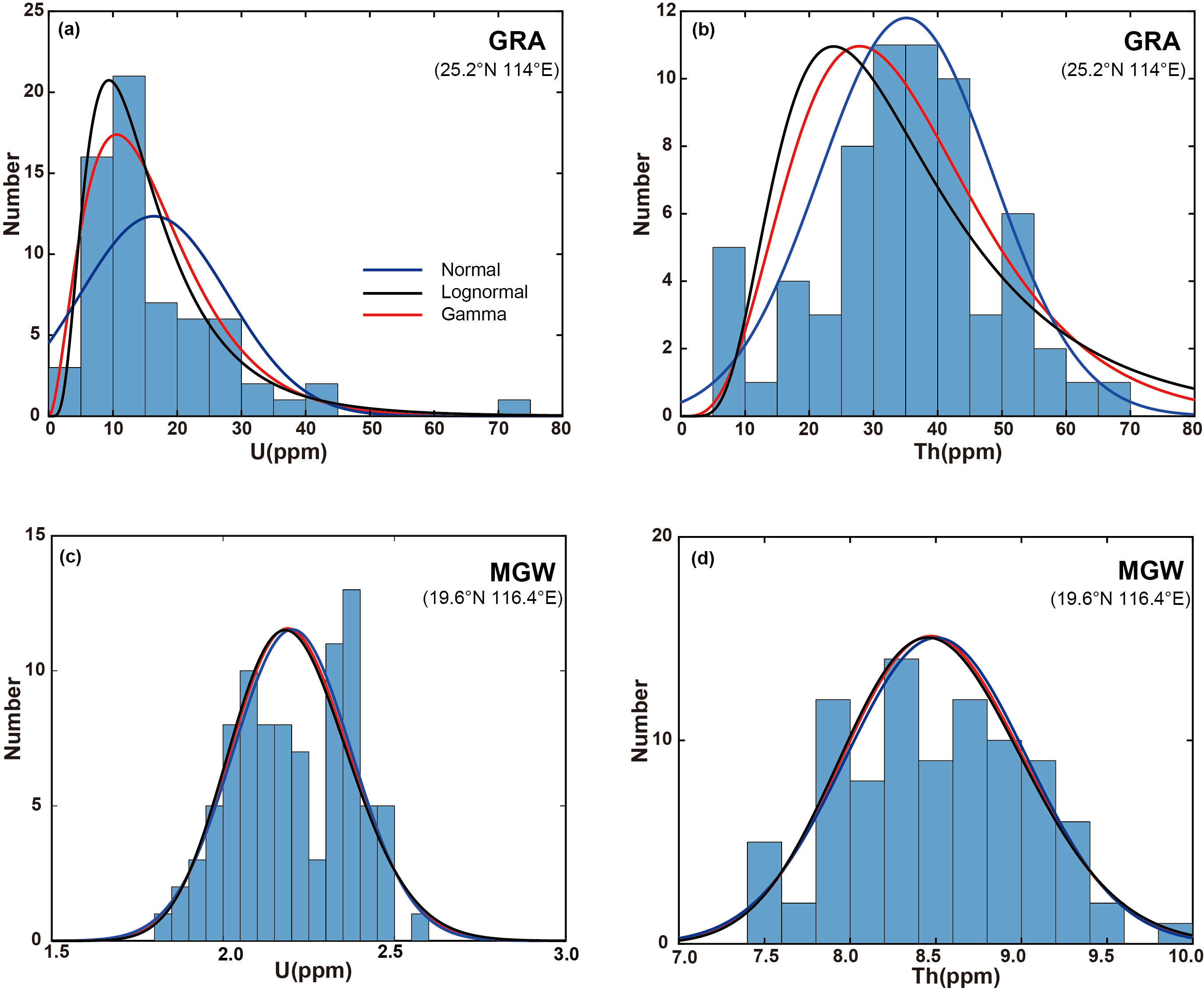}

\caption{The distribution of GRA and MGW \text{U/Th} with different fitting function.}
\label{fig:uthfitting}
\end{figure}

The fitting function selected depended on the shape of the curve for the abundance distribution in a particular rock type. When the shape was symmetric, i.e. the arithmetic mean and median are equal, normal distribution was selected. When the geometric mean was greater than the median, a log normal distribution was chosen. Takeuchi et.al(2019)\cite{TAKEUCHI2019} demonstrated that when a distribution is a highly skewed log normal distribution, the median may be smaller than the arithmetic mean to an order of magnitude, increasing the error in the fitted estimate. However, a gamma distribution does not cause a deviation from the mean, which would seem to render it a better choice than normal and lognormal distributions(see Fig.\ref{fig:uthfitting}).

When there were less than five samples of a certain rock type in a cell, the expected abundance and error were estimated using all the data for that rock type in the study area. This allowed for a more accurate estimate of the \text{U/Th} abundance in the cell, even when the sample size was small.
The 3-D probability density functions (PDFs) of \text{U/Th} concentrations in each cell were constructed by multiplying the abundance and posterior probability of each rock type in the cell, denoted as $A(i|x)$ and $P(i|x)$, respectively. 

By comparing the predicted and observed heat flow, the reliability of the 3-D \text{U} and \text{Th} concentrations model was verified. This allowed for a more accurate estimation of the \text{U/Th} concentrations in each cell, which could then be used to better understand the potential geoneutrino sources in the region. Overall, the integration of geological and heat data allowed for a more comprehensive characterization of the geological features in the region and provided valuable insights into the potential neutrino sources in the area.
%In the next step, it is planned to adjust the different depths of granite in the 3-D model to fit the heat flow data results. 

\section{Predicting the geoneutrino signal based on crustal model}

The joint geophysical and geochemical 3-D local crustal models developed around JUNO can calculate the geoneutrino flux produced by the Earth's crust. However, predicting the geoneutrino signal detectable at the JUNO detector requires information on particle detection via the inverse beta decay (IBD) reaction. The JULOC paper by R. Gao et al. \cite{JUNOgeoJULoc2020} outlines the detailed mathematics of the geoneutrino signal calculation. The detector efficiency $\epsilon(E)$ was assumed to be 1, and the cross-section $\sigma(E)$ was taken from Strumia and Vissani \cite{STRUMIA200342}. The geoneutrino energy spectrum $f_i(E)$ was taken from the website of ``Geoneutrino Spectrum and Luminosity''
 \cite{geospectrumurl}. Using these parameters and the 3-D local crustal models developed around JUNO, the calculated geoneutrino signal detectable at the JUNO detector.

The JULOC model used a simple approximation for the survival probability of electron antineutrinos. The approximation assumed the baseline-averaged probability to be $P_{ee} = 0.55$. However, this approximation may not be accurate enough given the improved experimental accuracy of the oscillation parameters and current and future geoneutrino measurements.

JULOC-I considered precision oscillation effects during the crustal geoneutrino calculations. The three-flavor survival probability for electron antineutrinos $\bar{\nu}_e$ propagating in vacuum is described as:

\begin{equation}
   P_{ee} = 1- P_0-P_\ast;
    \label{Eq:pee}
\end{equation}
with
\begin{equation}
   P_{0} = sin^22\theta_{12}cos^4\theta_{13}sin^2\Delta_{21},\\
      \label{Eq:p0}
\end{equation} 
\begin{equation}
   P_{\ast}=\frac{1}{2}sin^22\theta_{13}(1-cos\Delta_{\ast}cos\Delta_{21})\\
            +cos2\theta_{12}sin\Delta_{\ast}sin\Delta_{21},
    \label{Eq:p*}
\end{equation}

\begin{equation}
  \Delta_{ij}=1.27(\Delta m^2_{ij}L)/E_{\bar\nu},
  \Delta_{\ast} = \Delta_{31}+\Delta_{32}
    \label{Eq:delta}
\end{equation}
where $\theta_{12}$ and $\theta_{13}$ are the mixing angles, $\Delta m^2_{ij}$ is the square mass difference of antineutrinos in $eV^2$, $E_{\bar\nu}$ is the antineutrino energy in MeV, and L is the propagation distance in meters. We assume the mass ordering to be normal ($m1 < m2 < m3$), and the oscillation parameters are taken from the Particle Data Group (PDG). Finally, the density and abundance for each cell are taken from the new JULOC-I model.

To calculate the geoneutrino flux produced by the local crust (LOC) surrounding JUNO, the regional crust was divided into three layers, with each layer further divided into $0.4^{\circ}\times 0.4^{\circ} \times 1$ km cells. Each cell was assigned spatial, geophysical, and geochemical attributes. The total LOC contribution to the geoneutrino signal at JUNO was obtained by summing the signal of each cell, layer by layer (see Table \ref{tab:Loccontribution}).The JULOC-I integrated model updated the JULOC model only in terms of the contribution made by the continental crust. The contribution made by the oceanic crust was kept the same, as JULOC-I did not involve any new oceanic crust sampling.

To calculate the geoneutrino flux produced by the local crust (LOC) surrounding JUNO, the regional crust was divided into three layers, each with $0.4^{\circ}\times 0.4^{\circ} \times 1$ km cells, and assigned spatial, geophysical, and geochemical attributes. The total contribution of the LOC to the geoneutrino signal at JUNO was obtained by summing the signal of each cell layer by layer (see Table \ref{tab:Loccontribution}). The JULOC-I integrated model updated the contribution made by the continental crust of the JULOC model. The local continental crust was divided into two parts: the continental crust and the continental margin. The sum of these two parts provides the LOC continental contribution number in Table \ref{tab:Loccontribution}. The contribution of the oceanic crust remained unchanged, as JULOC-I did not involve new sampling of the oceanic crust.

\begin{figure}
\centering
\noindent\includegraphics[width=0.325\textwidth]{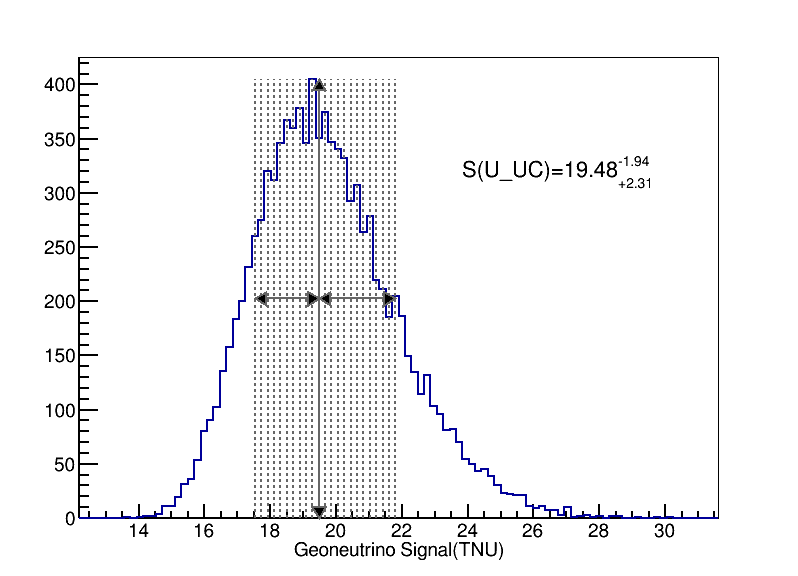}
\noindent\includegraphics[width=0.325\textwidth]{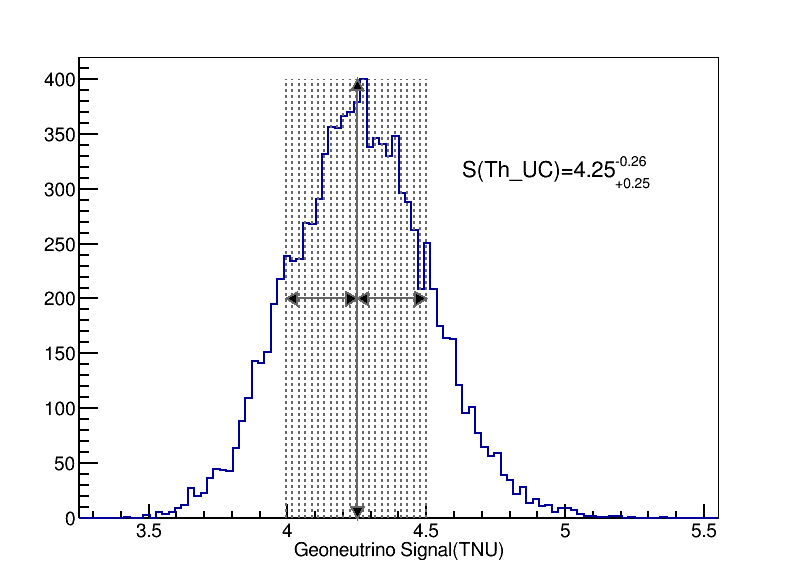}
\noindent\includegraphics[width=0.325\textwidth]{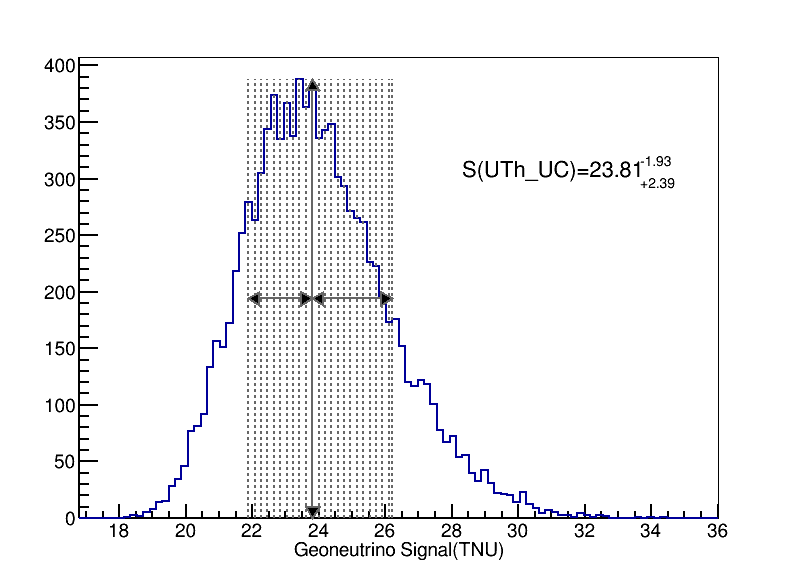}
\noindent\includegraphics[width=0.325\textwidth]{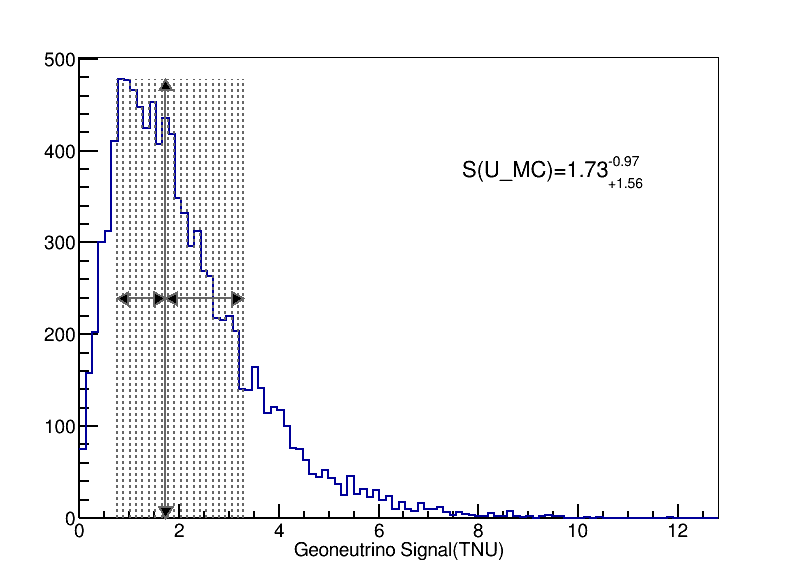}
\noindent\includegraphics[width=0.325\textwidth]{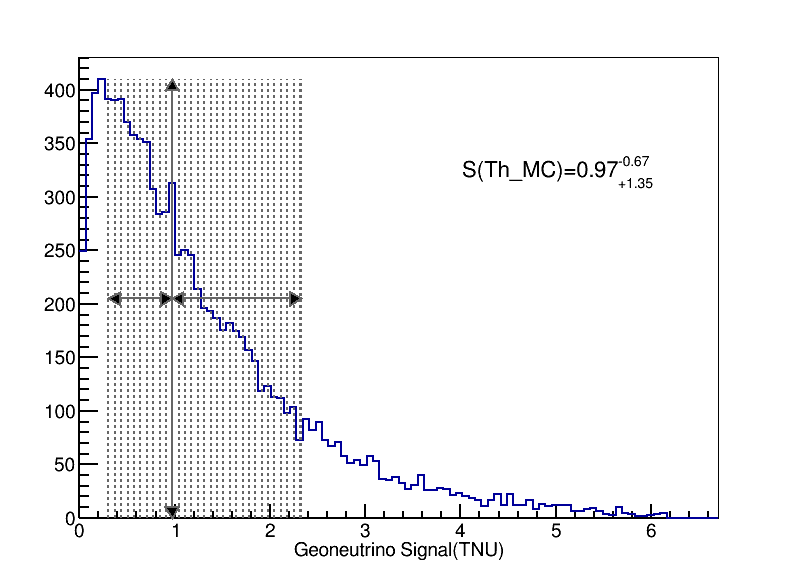}
\noindent\includegraphics[width=0.325\textwidth]{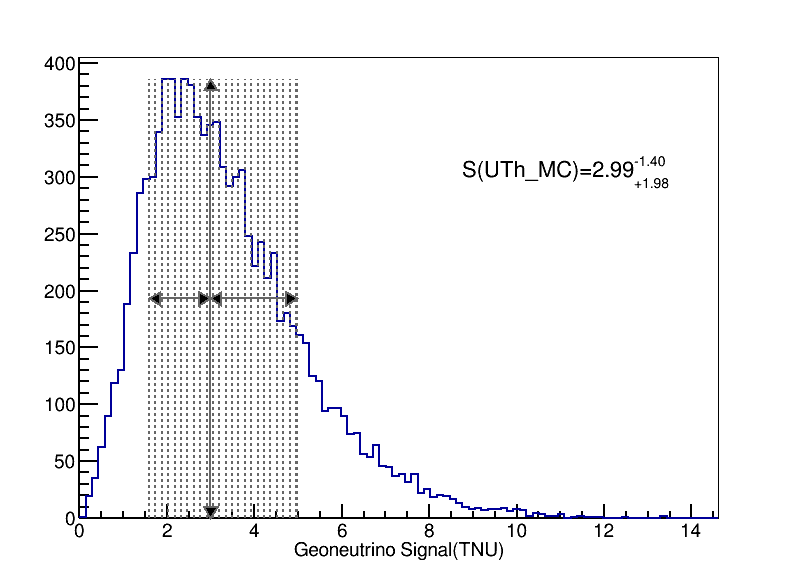}
\noindent\includegraphics[width=0.325\textwidth]{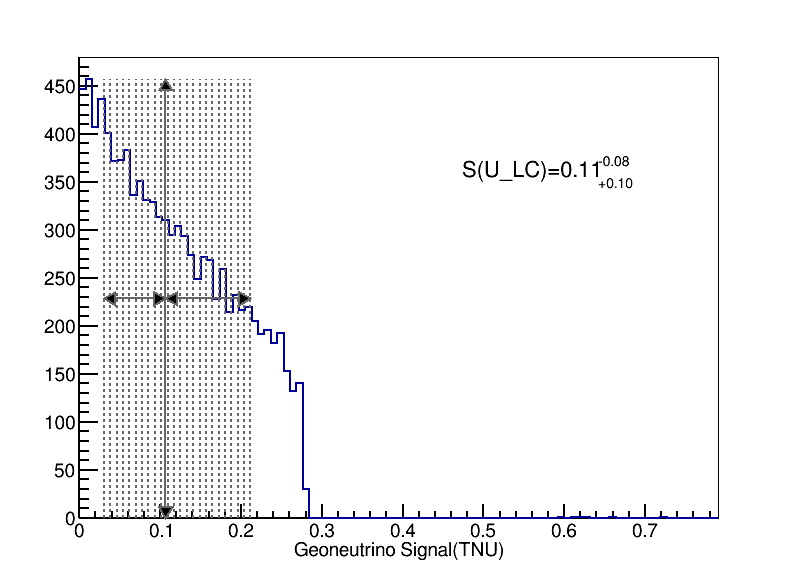}
\noindent\includegraphics[width=0.325\textwidth]{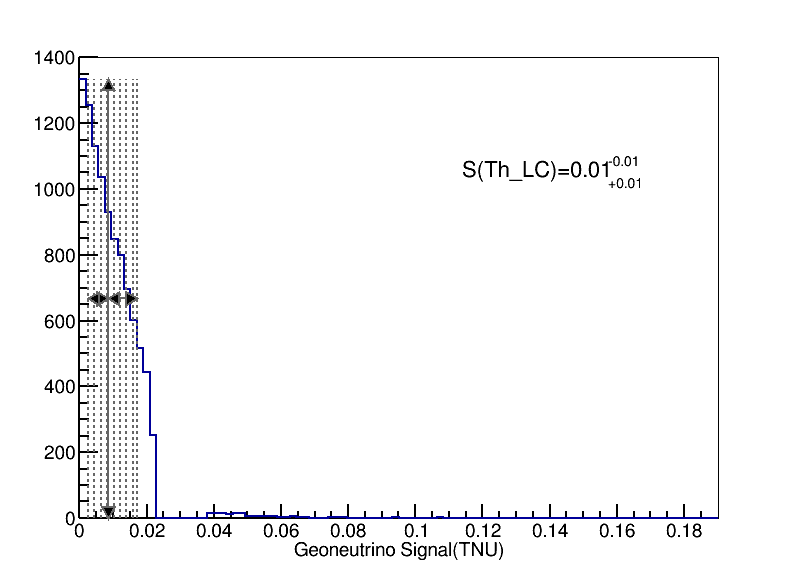}
\noindent\includegraphics[width=0.325\textwidth]{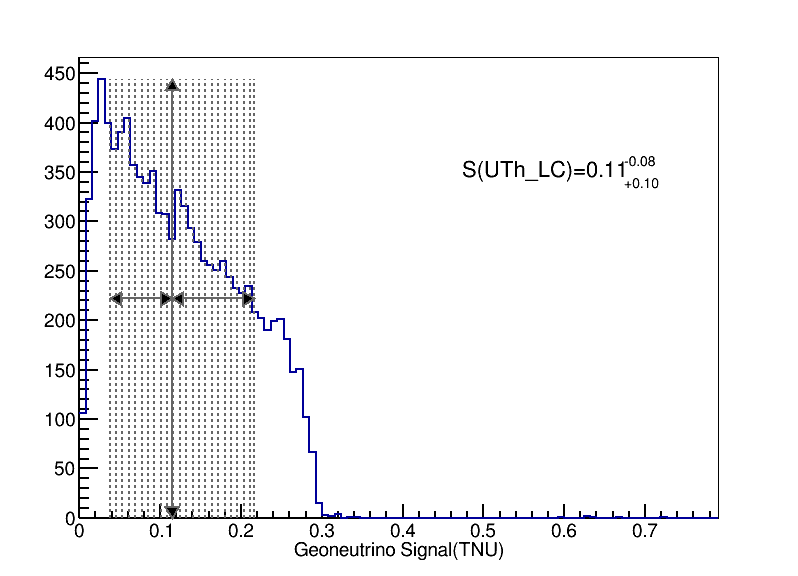}
\noindent\includegraphics[width=0.325\textwidth]{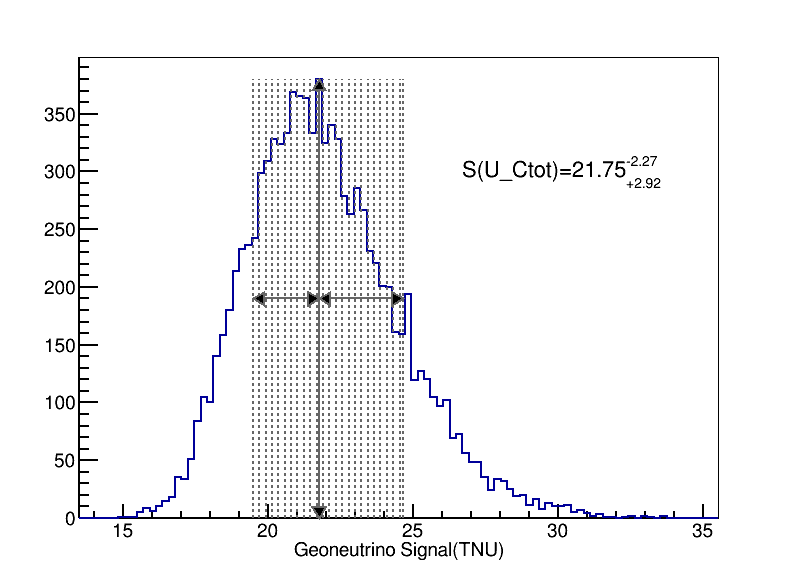}
\noindent\includegraphics[width=0.325\textwidth]{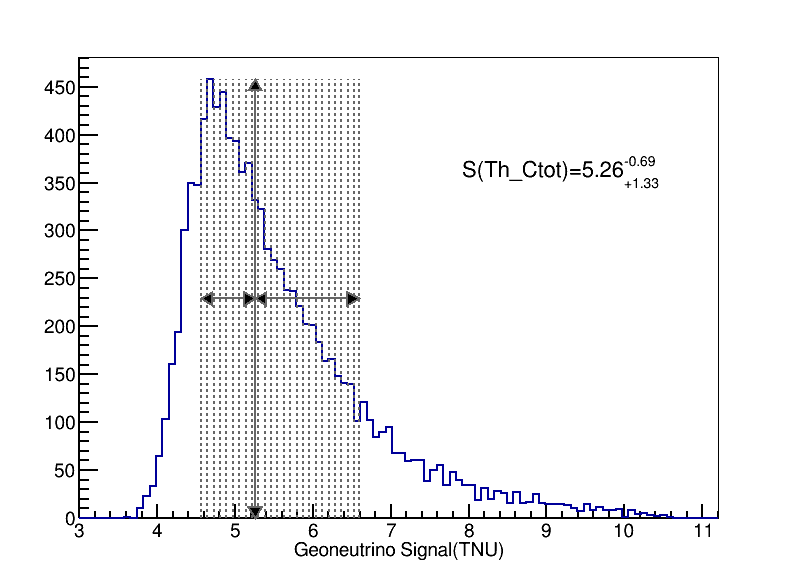}
\noindent\includegraphics[width=0.325\textwidth]{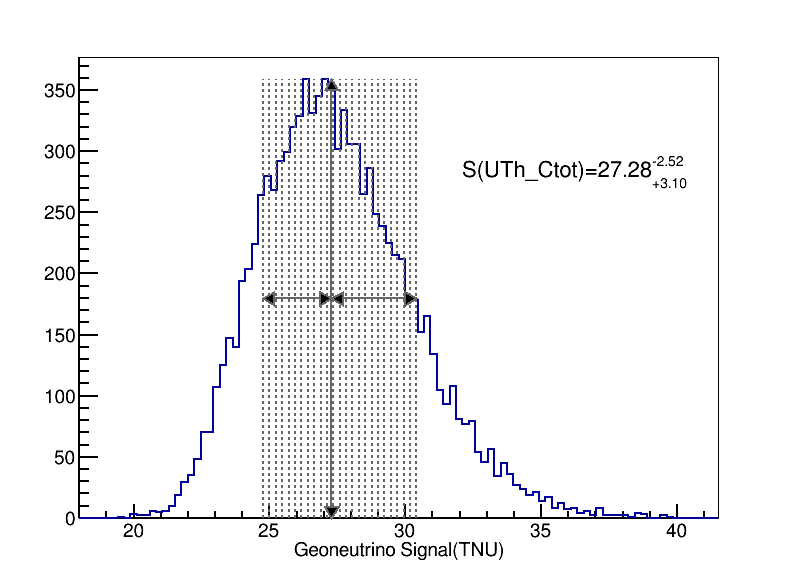}
\caption{The figure displays the distribution of the geoneutrino signal from
$^{238}$U (left), $^{232}$Th (middle), and their sum (right) for the UC (upper), MC
(upper middle), LC (lower middle), and the total (bottom). The median value
of the distribution represents the central value of the geoneutrino signal, and
the regions corresponding to 1 standard deviation (68.3\%) on the left and right
represent its upper and lower uncertainties, respectively. The results pertain
only to the continental crust, and neither continental margin.
nor the oceanic is included.}
\label{fig:1000dis}
\end{figure}

\begin{figure}
\centering
\noindent\includegraphics[width=0.325\textwidth]{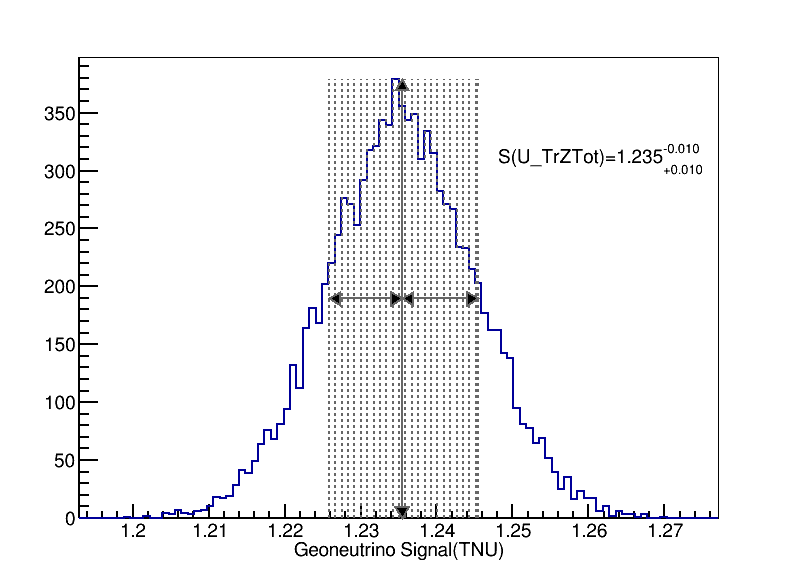}
\noindent\includegraphics[width=0.325\textwidth]{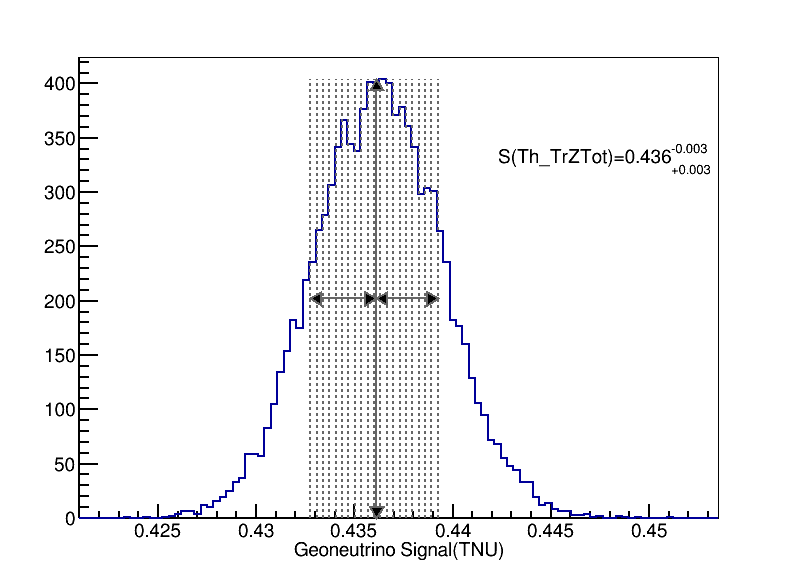}
\noindent\includegraphics[width=0.325\textwidth]{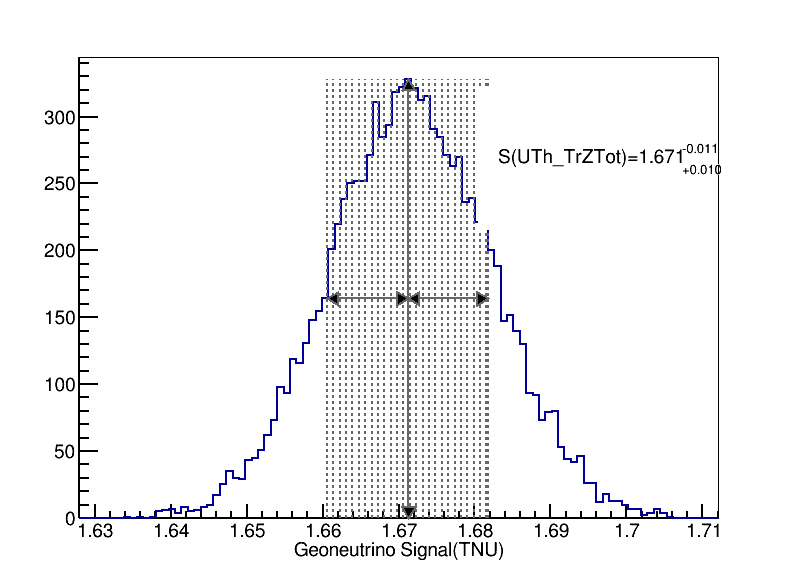}
\caption{ The figure shows the distribution of the geoneutrino signal emitted by $^{238}$U (left panel), $^{232}$Th (middle panel), and their combined contribution (right panel) within the continental margin. }
\label{fig:10000distanszone}
\end{figure}

Figure \ref{fig:1000dis} shows the distribution of the geoneutrino signal derived from 10,000 randomly selected samples. These samples were chosen based on the statistical distribution of input parameters, such as the density, layer thickness, and abundances of U and Th in each cell within the Continental Crust. The uncertainties associated with the input random numbers were propagated through the geoneutrino signal calculation, resulting in a distribution of the geoneutrino signal. We adopted the median value of the distribution as the central value to represent the geoneutrino signal, with the left and right regions corresponding to 1 standard deviation (68.3\%) serving as its upper and lower uncertainties, respectively. This method offers a comprehensive and precise representation of the geoneutrino signal distribution and its associated uncertainties.  Fig.\ref{fig:10000distanszone} illustrates the geoneutrino signal emitted by U and Th, as well as their combined contribution within the continental margin.

\end{appendices}
\section*{Data availability}
The data underlying this article will be shared on reasonable request to the corresponding author.

\bibliographystyle{apsrev4-1}
\bibliography{reference.bib}

\end{document}